\itshape
\documentclass[useAMS,usenatbib,usegraphicx]{mn2e}
\usepackage{graphicx}

\title[SN 2008ax]{The He-rich Stripped-Envelope Core-Collapse Supernova 2008ax.\thanks{This paper is based on observations collected at the 3.58m Telescopio Nazionale Galileo (La Palma, Spain), the 2.2m Telescope of the Centro Astron\'{o}mico Hispano Alem\'{a}n (Calar Alto, Spain), the Asiago 1.82m and 1.22m Telescopes (Italy) and the 1.08m AZT-24 telescope (Campo Imperatore, Italy).}}
\author[S.~Taubenberger et al.]{S.~Taubenberger$^{1}$\thanks{E-mail:
tauben@mpa-garching.mpg.de}, H.~Navasardyan$^{2}$, J.~I. Maurer$^{1}$, L.~Zampieri$^{2}$, N.~N. Chugai$^{3}$, 
\newauthor S.~Benetti$^{2}$, I.~Agnoletto$^{2,4}$, F.~Bufano$^{2}$, N.~Elias-Rosa$^{5,6}$, M.~Turatto$^{7,8}$, F.~Patat$^{9}$, 
\newauthor E.~Cappellaro$^{2}$, P.~A. Mazzali$^{1,2,10}$, T.~Iijima$^{2}$, S.~Valenti$^{2,11}$, A.~Harutyunyan$^{12}$, 
\newauthor R.~Claudi$^{2}$ and M.~Dolci$^{13}$ \\
$^{1}$Max-Planck-Institut f\"{u}r Astrophysik, Karl-Schwarzschild-Str. 1, 85741 Garching bei M\"{u}nchen, Germany\\
$^{2}$INAF Osservatorio Astronomico di Padova, Vicolo dell'Osservatorio 5, 35122 Padova, Italy\\
$^{3}$Institute of Astronomy, RAS, Pyatnitskaya 48, 119017 Moscow, Russia\\
$^{4}$Universit\`{a} degli studi di Padova, Vicolo dell'Osservatorio 3, 35122, Padova, Italy\\
$^{5}$Spitzer Science Center, California Institute of Technology, 1200 E. California Blvd, Pasadena, California 91125, USA\\
$^{6}$Institut d'Estudis Espacials de Catalunya, c/ Gran Capit\`a 24, 08034 Barcelona, Spain\\
$^{7}$INAF Osservatorio Astrofisico di Catania, Via S.Sofia 78, 95123 Catania, Italy\\
$^{8}$INAF Osservatorio Astronomico di Trieste, Via Tiepolo 11, 34143 Trieste, Italy\\ 
$^{9}$European Organisation for Astronomical Research in the Southern Hemisphere (ESO), Karl-Schwarzschild-Str. 2, \\
~\,85748 Garching bei M\"{u}nchen, Germany\\
$^{10}$Scuola Normale Superiore, Piazza dei Cavalieri 7, 56126 Pisa, Italy\\
$^{11}$Astrophysics Research Centre, School of Mathematics and Physics, Queen's University Belfast, Belfast BT7 1NN, UK\\ 
$^{12}$Fundaci\'{o}n Galileo Galilei-INAF, Telescopio Nazionale Galileo, E-38700 Santa Cruz de la Palma, Tenerife, Spain \\
$^{13}$INAF Osservatorio Astronomico di Collurania, via M. Maggini, I-64100 Teramo, Italy\\}

\begin{document}

\date{Accepted 2011 January 4. Received 2010 December 28; in original form 2010 April 12}

\pagerange{\pageref{firstpage}--\pageref{lastpage}} \pubyear{2011}

\maketitle

\label{firstpage}

\begin{abstract}
Extensive optical and near-infrared (NIR) observations of the type IIb supernova 2008ax are presented, covering the first year after the explosion. The light curve is mostly similar in shape to that of the prototypical type IIb SN~1993J, but shows a slightly faster decline rate at late phases and lacks the prominent narrow early-time peak of SN~1993J. From the bolometric light curve and ejecta expansion velocities, we estimate that about 0.07--0.15\ $M_\odot$ of $^{56}$Ni were produced during the explosion and that the total ejecta mass was between 2 and 5 $M_\odot$, with a kinetic energy of at least $10^{51}$ erg. The spectral evolution of SN~2008ax is similar to that of the type Ib SN~2007Y, exhibiting high-velocity Ca II features at early phases and signs of ejecta-wind interaction from H$\alpha$ observations at late times. NIR spectra show strong He I lines similar to the type Ib SN~1999ex, and a large number of emission features at late times. Particularly interesting are the strong, double-peaked He I lines in late NIR spectra, which -- together with double-peaked [O I] emission in late optical spectra -- provide clues for asymmetry and large-scale Ni mixing in the ejecta.
 
\end{abstract}

\begin{keywords}
supernovae: general -- supernovae: individual (SN~2008ax, SN~1993J, SN~1999ex, SN~2007Y) -- galaxies: individual (NGC~4490).
\end{keywords}

\section{Introduction}
Type IIb, Ib and Ic supernovae (SNe~IIb, Ib, Ic), also referred to as stripped-envelope supernovae (SE-SNe), show a large diversity in observed properties. They are distinguished from the majority of all core-collapse events (type IIP SNe) by less persistent or absent hydrogen lines in their early spectra and the lack of a plateau phase in their light curves. SNe~IIb represent a transition between hydrogen-dominated SNe~IIP and mostly hydrogen-free, helium-dominated SNe~Ib. They are therefore particularly interesting to clarify the evolutionary path of SE-SN progenitors. However, SNe IIb are relatively rare. Only few have been observed up to now, and little is known about their spread in mass and energy. 

The subdivision of SNe into type IIb and type Ib is not uncompromising, as it depends strongly on the hydrogen mass and distribution, and on the phase at which the supernova was discovered.  Recent studies have suggested that there may be traces of hydrogen in all SN~Ib spectra, and of both hydrogen and helium even in SNe~Ic \citep{br02,br06,e2,par07}. 

Despite the increasing interest in SE-SNe due to their link to long-duration $\gamma$-ray bursts \citep[GRBs; e.g.][]{galama98,hjorth03,mirabal06,campana06} and X-ray flashes (\citealt{sollerman06,maz08}; but see also e.g. \citealt{mod09}), their diversity has not yet been fully understood. New observations, like the first detection of a very energetic type IIb hypernova (SN~2003bg; \citealt{hamuy09,mazzali09}), rather indicate more diversity than previously thought. At present, possible progenitor scenarios include massive stars in close binary systems experiencing strong mass transfer, very massive single stars with strong stellar winds \citep{heg03}, or a combination of these two processes. So far, it has not been possible to discriminate between these scenarios by direct progenitor detections. 
SN~2008ax provides a rare opportunity to add detailed observations of a SN of this class. It is one of few cases discovered soon after explosion, with the progenitor detected in HST images \citep{l1,c1}. From X-ray to radio wavelengths, it is the best-monitored SN~IIb after the very nearby SN~1993J \citep[e.g.][]{vandriel93,lewis94,o94,ba1,richmond96}, which has become a prototype for the entire subclass. 

SN~2008ax was discovered independently by \citet{m1} and \citet{n1}. With coordinates $\alpha = 12^{\rmn{h}} 30^{\rmn{m}} 40\fs8$ and $\delta = +41\degr 38\arcmin 14\arcsec.5$ (J2000) it was located $53.1$ arcsec east and $25.8$ arcsec south of the centre of NGC~4490. Follow-up observations of SN~2008ax started soon after discovery. The SN site had been monitored by \citet{a1} on UT March 3.19, and no source had been detected down to a limiting magnitude of 18.5. This detection limit allowed \citet[][hereafter P08]{p1} to constrain the explosion time to be UT March 3.30, with a small uncertainty of 0.15 days (JD = $2\,454\,528.80 \pm  0.15$). Early X-ray, UV, optical and radio observations of SN~2008ax were presented by \citet[][hereafter R09]{r09}, and densely sampled optical photometry was published by P08 and \citet[][hereafter Ts09]{ts09}. A series of optical and early NIR spectra were presented by P08 and \citet[][hereafter Ch10]{Ch10}. Three epochs of optical spectropolarimetry were analysed by Ch10, and reveal strong line polarisation, especially across the hydrogen lines, shortly after the explosion.
23 GHz Very-Long-Baseline-Interferometry (VLBI) observations of SN~2008ax, made 33 days after the explosion with the Very Long Baseline Array (VLBA) by \citet{m3}, resulted in a marginal detection of the supernova. A total flux density of $0.8\pm0.3$ mJy was measured, and the structure was interpreted as either a core-jet or a double source.  A progenitor mass-loss rate of $\dot M  = (9\pm3)\times10^{-6}\ M_\odot$\,yr$^{-1}$ was inferred by R09 based on X-ray data. Ch10 estimated $\dot M  \leq 10^{-5}\ M_\odot$\,yr$^{-1}$ by the lack of narrow emission lines in a moderate-resolution early-time spectrum, in good agreement with the R09 result. R09 detected an initial fading in the UV light curves of SN~2008ax, interpreted as a fingerprint of adiabatic cooling after shock breakout.
With this paper we contribute additional broadband optical and infrared imaging and spectroscopy of SN~2008ax, starting shortly after the explosion and extending to one year thereafter. Modelling of our nebular spectra has already been published by \citet[][hereafter M10]{M10}.

\section{Observations and Data Reduction}
\subsection{Photometry}
Optical ($U\!BV\!RI$) imaging of SN~2008ax was carried out using the following instruments: 

\begin{enumerate}
\item DOLORES (with a scale of 0.252 arcsec px$^{-1}$ and a field of view of $8.6\times8.6$ arcmin$^{2}$) at the 3.58m Telescopio Nazionale Galileo (La Palma, Spain)

\item CAFOS (with a scale of 0.53 arcsec px$^{-1}$ and a field of view of $9\times9$ arcmin$^{2}$) at the Calar Alto 2.2m Telescope (Andaluc\'{i}a, Spain)

\item AFOSC (with a scale of 0.46 arcsec px$^{-1}$ and a field of view of $7.8\times7.8$ arcmin$^{2}$) mounted on the 1.82m Copernico Telescope of Mt. Ekar (Asiago, Italy)
\end{enumerate}

\begin{figure}
	\includegraphics[width=8.4cm]{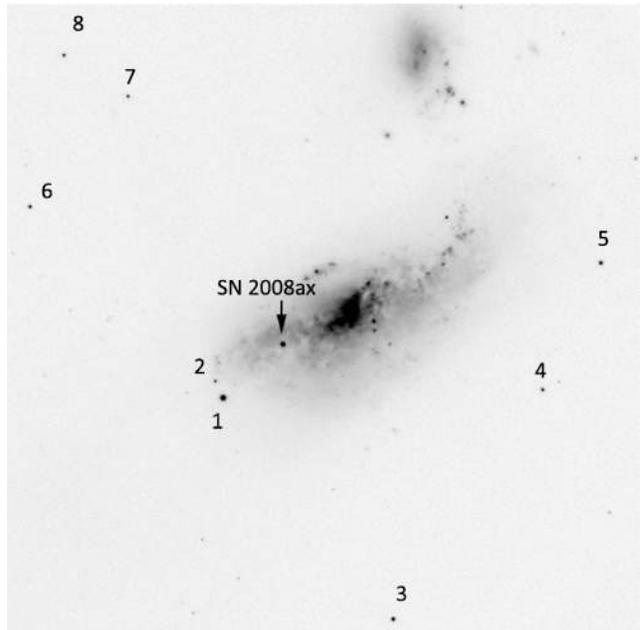}
	\caption{$V$-band image of SN~2008ax in NGC~4490, taken with CAFOS on March 8. The field of view is about $8\times8$ arcmin$^2$, north is up and east to the left. Photometric comparison stars are labelled.}
	\label{fig:sn08axseq}
\end{figure}

The optical data were reduced following standard prescriptions (overscan, bias, flat-field correction) in the \textsc{\textsc{iraf}}\footnote{Image Reduction and Analysis Facility (\textsc{\textsc{iraf}}), a software system distributed by the National Optical Astronomy Observatories (NOAO).} environment. Instrumental magnitudes were measured on the reduced images using a PSF-fitting technique with the software package \textsc{\textsc{snoopy}}, specifically developed for SN photometry by E. Cappellaro and F. Patat. For late-time $B$- and $V$-band data we used the template-subtraction technique with pre-explosion reference images of the host galaxy. Several tests performed with the PSF-fitting and template-subtraction techniques showed negligible differences as long as the SN was brighter than $\sim$\,19 mag.

The calibration of the optical photometry was performed with respect to standard fields of \citet{la92}, observed in the same nights as the SN. The magnitudes of a sequence of stars in the field of NGC~4490 (Fig.~\ref{fig:sn08axseq} and Table~\ref{seq_mags}) were computed averaging the measurements obtained during four photometric nights, with the uncertainties reported in Table~\ref{seq_mags} being the r.m.s. deviations over these nights. Care was taken to exclude saturated stars from the analysis. The SN magnitudes were finally determined relative to these stars. Since instruments with different passbands were used for the follow-up of SN~2008ax (see fig.~2 of \citealt{tau10}), we applied the `$S$-correction' technique \citep{s2,pig04} to calibrate the optical SN magnitudes to the standard photometric system of Johnson and Cousins \citep{bes90}. The $S$-correction was determined from our spectral sequence of SN~2008ax. At epochs where no spectra were available, the $S$ terms were determined by linear interpolation or constant extrapolation. 

NIR ($JHK$) imaging was obtained with NICS (0.25 arcsec px$^{-1}$, field of view $4.2\times4.2$ arcmin$^{2}$) mounted on the 3.58m Telescopio Nazionale Galileo (La Palma, Spain) and SWIRCAM (1.03 arcsec px$^{-1}$, field of view $4.4\times4.4$ arcmin$^{2}$) mounted on the 1.08m AZT-24 Telescope (Campo Imperatore, Italy). 

The pre-reduction of the NIR images required some additional steps. Owing to the bright background in the NIR, we had to remove the sky contribution from the target images. This was done by median-combining a number of dithered science frames. The resulting sky template was then subtracted from the target images. Our data were obtained in several dithered, short exposures to enable a clean sky subtraction and to avoid exceeding the linearity regimes of the detectors. The NICS images were additionally corrected for cross-talk and for the distortion of the NICS optics. These corrections were performed using the SNAP\footnote{Written by F. Mannucci,\\
http:/$\!$/www.tng.iac.es/news/2002/09/10/snap/} pipeline available at TNG for the reduction of NICS data.

In analogy to the optical observations, NIR instrumental magnitudes were measured using a PSF-fitting technique. Only two stars of our local sequence (stars 1 and 2 in Fig.~\ref{fig:sn08axseq} and Table~\ref{seq_mags}) were contained in the field of view of the NIR instruments. They were calibrated on three photometric nights using standard fields of the Arnica catalogue \citep{h}. Again, the SN photometry was derived relative to the local standards. 

\begin{figure}
   \centering
   \includegraphics[width=8.4cm]{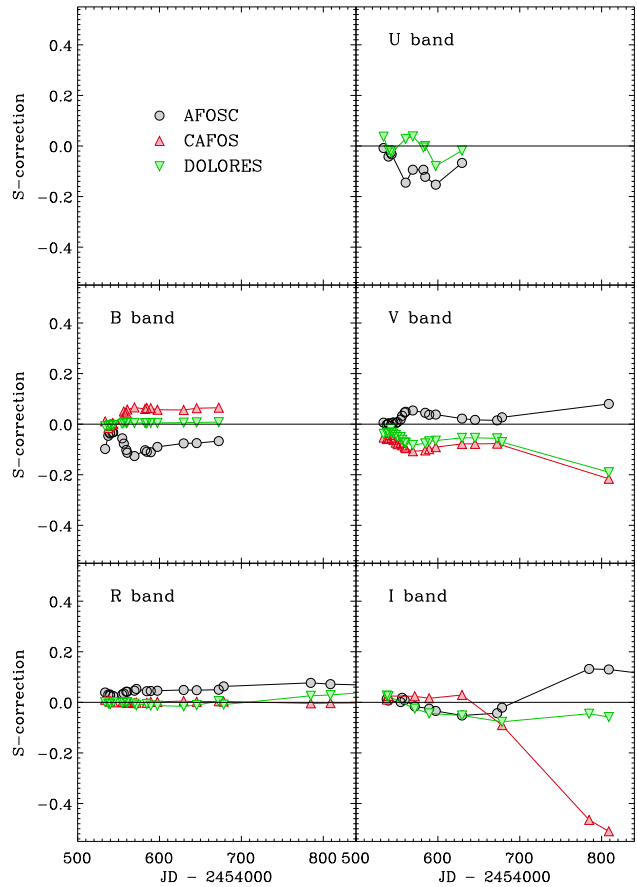}
   \caption{Temporal evolution of the $S$-correction in the optical bands 
   for the different instrumental configurations used for follow-up observations of SN~2008ax.}
   \label{fig:Scorr}
\end{figure}

The calibrated SN magnitudes in optical and infrared bands are reported in Tables~\ref{opt_mags} and \ref{IR_mags}, respectively. The $S$-correction for the optical bands (i.e., the quantity added to the zero-point calibrated magnitudes instead of a colour-term correction) is listed in Table~\ref{Scorr}, and its temporal evolution for the different instruments is shown in Fig.~\ref{fig:Scorr}.

\begin{table*}
 \centering
 \begin{minipage}{175mm}
  \caption{$S$-corrected optical photometry of SN~2008ax.}
  \label{opt_mags}
  \begin{tabular}{@{}lcrccccccl@{}}
  \hline
   Date & JD & Phase\footnote{Phase in days with respect to the explosion date (JD = $2\,454\,528.80\pm0.15$). $B$-band maximum light occurred on day 18.3.} &$U$ & $B$ & $V$ & $R$ & $I$ & Seeing\footnote{Average seeing in arcsec over all filter bands.} & Instr.\footnote{CAFOS = Calar Alto 2.2m Telescope + CAFOS; DOLORES = 3.58m Telescopio Nazionale Galileo + DOLORES; AFOSC = Asiago 1.82m Copernico Telescope + AFOSC.} \\
   & $-2\,454\,000$ & (days) & & & & & & (arcsec) & \\

\hline
08/03/08 & 533.59&  4.8 &               &$16.50\pm0.03$&$15.60\pm0.03$&$15.13\pm0.03$&$14.85\pm0.04$& 1.5 & CAFOS\\
12/03/08 & 537.53&  8.7 &               &$15.30\pm0.04$&$14.62\pm0.02$&$14.35\pm0.02$&$14.07\pm0.05$& 2.1 & AFOSC\\
14/03/08 & 539.76& 10.9 & $14.47\pm0.10$&$14.86\pm0.07$&$14.20\pm0.03$&$13.96\pm0.05$&$13.71\pm0.03$& 0.9 & DOLORES\\
16/03/08 & 542.41& 13.6 &               &$14.47\pm0.04$&$13.91\pm0.03$&$13.66\pm0.02$&$13.36\pm0.06$& 1.4 & CAFOS\\
28/03/08 & 554.47& 25.7 & $14.34\pm0.21$&              &$13.72\pm0.05$&$13.32\pm0.13$&$12.97\pm0.17$& 2.8 & AFOSC\\
30/03/08 & 556.44& 27.6 & $15.35\pm0.11$&$14.96\pm0.06$&$13.92\pm0.03$&$13.45\pm0.02$&$13.02\pm0.02$& 1.3 & AFOSC\\
03/04/08 & 559.50& 30.7 & $16.27\pm0.12$&$15.44\pm0.05$&$14.19\pm0.03$&$13.61\pm0.04$&$13.17\pm0.01$& 3.1 & AFOSC\\
04/04/08 & 560.61& 31.8 &               &$15.54\pm0.03$&$14.23\pm0.04$&$13.61\pm0.02$&$13.15\pm0.04$& 2.1 & CAFOS\\
05/04/08 & 561.56& 32.8 &               &$15.65\pm0.04$&$14.30\pm0.03$&$13.66\pm0.02$&$13.18\pm0.05$& 1.9 & CAFOS\\
12/04/08 & 569.35& 40.6 &               &$16.10\pm0.03$&$14.69\pm0.05$&$14.06\pm0.02$&$13.40\pm0.04$& 1.5 & CAFOS\\
15/04/08 & 571.75& 42.9 &               &$16.25\pm0.04$&$14.81\pm0.12$&$14.18\pm0.06$&$13.48\pm0.02$& 2.3 & DOLORES\\
25/04/08 & 582.35& 53.6 &               &$16.38\pm0.04$&$15.03\pm0.03$&$14.44\pm0.03$&$13.71\pm0.05$& 1.2 & CAFOS\\
01/05/08 & 588.47& 59.7 &               &$16.44\pm0.03$&$15.16\pm0.04$&$14.57\pm0.03$&$13.80\pm0.03$& 1.1 & CAFOS\\
02/05/08 & 589.38& 60.6 &               &$16.47\pm0.04$&$15.22\pm0.03$&$14.69\pm0.01$&$13.95\pm0.04$& 1.7 & AFOSC\\
10/05/08 & 597.39& 68.6 & $17.08\pm0.11$&$16.67\pm0.03$&$15.38\pm0.02$&$14.82\pm0.01$&$13.96\pm0.01$& 1.0 & DOLORES\\
11/06/08 & 629.48& 100.7 &$17.35\pm0.05$&$16.99\pm0.06$&$16.01\pm0.05$&$15.49\pm0.04$&$14.62\pm0.04$& 1.2 & DOLORES\\
27/06/08 & 645.37& 116.6 &              &$17.32\pm0.07$&$16.37\pm0.05$&$15.82\pm0.02$&$14.85\pm0.05$& 1.4 & CAFOS\\
05/07/08 & 653.41& 124.6 &              &$17.41\pm0.04$&$16.50\pm0.05$&$15.97\pm0.03$&$14.99\pm0.03$& 1.8 & CAFOS\\ 
24/07/08 & 672.39& 143.6 &$18.10\pm0.10$&$17.82\pm0.03$&$17.08\pm0.05$&$16.40\pm0.02$&$15.46\pm0.02$& 2.2 & AFOSC\\
24/11/08 & 794.69& 265.9 &              &$19.86\pm0.16$&$19.24\pm0.16$&$18.33\pm0.02$&$18.02\pm0.07$& 1.6 & DOLORES\\
22/12/08 & 822.76& 294.0 &              &$20.40\pm0.13$&$19.67\pm0.04$&$18.67\pm0.02$&$18.42\pm0.05$& 0.9 & DOLORES\\
19/02/09 & 881.54& 352.7 &              &$21.34\pm0.19$&$20.52\pm0.12$&$19.59\pm0.10$&$19.16\pm0.10$& 1.2 & CAFOS\\
\hline
\end{tabular}
\end{minipage}
\end{table*}

\begin{table*}
 \centering
 \begin{minipage}{115mm}
  \caption{NIR photometry of SN~2008ax.}
  \label{IR_mags}
  \begin{tabular}{@{}lcrcccl@{}}
  \hline
    Date & JD & Phase\footnote{Epoch in days with respect to the explosion date (JD = $2\,454\,528.80 \pm 0.15$). $B$-band maximum light occurred on day 18.3.}  & $J$ & $H$ & $K$ & Instr.\footnote{NICS = 3.58m Telescopio Nazionale Galileo + NICS; CI = Campo Imperatore 1.08m AZT-24 Telescope + SWIRCAM.} \\
    & $-2\,454\,000$ & (days) & & & & \\

 \hline
 13/03/08  & 539.61 & 10.8 & $13.41\pm0.04$ & $ 13.15\pm0.02$&$ 12.95\pm0.06$& NICS\\
 20/03/08  & 546.57 & 17.8 & $12.72\pm0.05$ & $ 12.55\pm0.06$&$ 12.35\pm0.06$& CI\\
 28/03/08  & 554.56 & 25.8 & $12.43\pm0.05$ & $ 12.23\pm0.05$&$ 12.01\pm0.05$& CI\\
 29/03/08  & 555.48 & 26.7 & $12.50\pm0.05$ & $ 12.23\pm0.05$&$ 12.05\pm0.06$& CI\\
 31/03/08  & 557.48 & 28.7 & $12.49\pm0.05$ & $ 12.26\pm0.05$&$ 12.05\pm0.05$& CI\\
 01/04/08  & 558.44 & 29.6 & $12.55\pm0.05$ & $ 12.29\pm0.05$&$ 12.09\pm0.05$& CI\\
 07/04/08  & 564.38 & 35.6 & $12.74\pm0.05$ & $ 12.41\pm0.06$&$ 12.22\pm0.06$& CI\\
 12/04/08  & 569.49 & 40.7 & $12.95\pm0.05$ & $ 12.61\pm0.06$&$ 12.48\pm0.06$& CI\\
 14/04/08  & 571.48 & 42.7 & $13.01\pm0.03$ & $ 12.66\pm0.04$&$ 12.53\pm0.07$& NICS\\
 11/06/08  & 629.54 & 100.7 & $15.17\pm0.03$ & $ 14.46\pm0.02$&$ 14.27\pm0.01$& NICS\\
 11/07/08  & 659.42 & 130.6 & $16.03\pm0.02$ & $ 15.30\pm0.02$&$ 15.18\pm0.03$& NICS\\
\hline
\end{tabular}
\end{minipage}
\end{table*}

\subsection{Spectroscopy}

The log of spectroscopic observations is reported in Table~\ref{spectra}. The instruments were mostly the same as for the photometry. Three additional spectra near maximum brightness were taken with the B\&C spectrograph mounted on the 1.22m Telescope in Asiago. We collected 27 epochs of spectroscopy, from 4.8 to 358.7 days after the explosion. To avoid flux losses due to differential refraction \citep{fil}, the slit was aligned at the parallactic angle when the object was observed at airmass higher than 1.3.

The spectra were reduced using \textsc{\textsc{iraf}} routines. Frames were de-biased and flat-field corrected before the extraction of the spectra. Wavelength calibration was accomplished with the help of comparison lamps. Flux calibration and telluric-line removal were done using spectrophotometric standard star spectra observed during the same nights. If necessary, the spectral fluxes were adjusted to match the photometry.

Four low-resolution NIR spectra of SN~2008ax were obtained with NICS at TNG (Table~\ref{spectra}). The spectral range of 8800--24500\,\AA\ was covered using the $IJ$ and $HK$ dispersers with resolving power 500. All NIR spectroscopic observations were split into sub-exposures taken at different positions along the slit following an ABBA scheme.
After dark and flat-field corrections and a pairwise subtraction of dithered frames, the SN spectra were optimally extracted, scaled to match in intensity and then combined. As for the optical spectroscopy, wavelength calibration was accomplished using arc lamps. In order to remove atmospheric absorption features and instrumental response, the object spectra were divided by a telluric standard of spectral type A0V, and re-scaled using the spectral energy distribution of Vega. Finally, the flux scale was adjusted to match the broadband $JHK$ magnitudes.

\begin{table*}
 \centering
 \begin{minipage}{118mm}
  \caption{Spectroscopic observations of SN~2008ax.}
  \label{spectra}
  \begin{tabular}{@{}lcrccc@{}}
  \hline
  Date & JD & Phase\footnote{Phase in days with respect to the explosion date (JD = $2\,454\,528.80\pm0.15$). $B$-band maximum light occurred on day 18.3.}  &Instrument\footnote{CAFOS = Calar Alto 2.2m Telescope + CAFOS; DOLORES = 3.58m Telescopio Nazionale Galileo + DOLORES; AFOSC = Asiago 1.82m Copernico Telescope + AFOSC; B\&C = Asiago 1.22m Galilei Telescope + B\&C; NICS = 3.58m Telescopio Nazionale Galileo + NICS.} & Range & Resolution\footnote{Measured from the full-width at half maximum of night-sky lines.} \\
 & $-2\,454\,000$ & (days) & configuration & (\AA) &  (\AA) \\

 \hline
08/03/08  & 533.59 & 4.8  & CAFOS+B200  & 3600--8800 & 14 \\
11/03/08  & 537.46 & 8.7  & AFOSC+gm2,4 & 3600--9500 & 24,38\\
13/03/08  & 539.49 & 10.7 & AFOSC+gm2,4 & 3600--9700 & 24,38\\
14/03/08  & 539.69 & 10.8 & NICS+IJ,HK  & 8800--24500 & 22,26\\
14/03/08  & 539.68 & 10.9 & DOLORES+LRB,LRR &3200--9300 & 11\\
16/03/08  & 542.42 & 13.6 & CAFOS+B200 & 3400--6400 & 14\\
18/03/08  & 543.53 & 14.7 & CAFOS+B200 & 3300--8850 & 13\\
20/03/08  & 545.54 & 16.7 & B\&C+600ll/mm &  4750--7160  & 5\\
23/03/08  & 548.60 & 19.8 & B\&C+600ll/mm &  4750--7160  & 5\\
25/03/08  & 550.60 & 21.8 & B\&C+600ll/mm &  4730--7150  & 5\\
28/03/08  & 554.50 & 25.7 & AFOSC+gm2,4  &   3600--9200  & 24,38\\
30/03/08  & 556.37 & 27.6 & AFOSC+gm2,4  &   3600--9400  & 24,38\\
02/04/08  & 559.45 & 30.6 & AFOSC+gm2,4  &   3600--9400  & 24,38\\
04/04/08  & 560.62 & 31.8 & CAFOS+ B200  &   3300--8850  & 14\\
12/04/08  & 569.38 & 40.6 & CAFOS+B200   &   3300--8850  & 13\\
14/04/08  & 571.49 & 42.7 & NICS+IJ,HK   &   8800--24500 & 22,26\\
14/04/08  & 571.76 & 43.0 & DOLORES+LR-R &   5200--9200 & 11\\
25/04/08  & 582.37 & 53.6 & CAFOS+B200   &   3300--6400  & 13\\
28/04/08  & 584.52 & 55.7 & CAFOS+B200   &   3500--8800 & 13\\
02/05/08  & 589.34 & 60.5 & AFOSC+gm2,4 &  3600--9400  & 24,38\\
10/05/08  & 597.41 & 68.6 & DOLORES+LR-B,LR-R &  3350--9300 & 11\\
11/06/08  & 629.49 & 100.7 & DOLORES+LR-B,LR-R &  3350--9300 & 11\\
12/06/08  & 629.56 & 100.8 & NICS+IJ,HK &   8800--24500 & 22,26\\
27/06/08  & 645.38 & 116.6 & CAFOS+B200 &   3300--6400  & 13\\
11/07/08  & 659.44 & 130.6 & NICS+IJ,HK & 8800--24500 & 22,26\\
24/07/08  & 672.34 & 143.6 & AFOSC+gm2,4 &  3600--9400  & 24,38\\
30/07/08  & 678.36 & 148.6 & CAFOS+G200   &  4850--10000 & 14\\
24/11/08  & 784.75 & 255.9 & DOLORES+LR-R & 5200--9200  & 15\\
08/12/08  & 808.77 & 280.0 & DOLORES+LR-R & 5200--9200  & 11\\
09/12/08  & 809.67 & 280.9 & AFOSC+gm4   & 4200--7700  & 24\\
25/02/09  & 887.53 & 358.7 & DOLORES+LR-R & 5200--9200  & 16\\
\hline
\end{tabular}
\end{minipage}
\end{table*}

\section{Distance and reddening}

NGC~4490 is the brighter member of a closely interacting pair of galaxies, variously classified as a spiral or irregular system. In RC3 \citep{rc3} NGC~4490 is classified as type SB(s)d, and the smaller companion NGC~4485 as type IB(s)m. Both galaxies show signs of tidal disruption (see Fig.~\ref{fig:sn08axseq}). NGC~4485/90 is well studied over a wide range of wavelengths (from radio to X-rays). Its properties have already been discussed by P08 and \citet{c1}. SN~1982F, classified as SN~IIP, was also detected in the same galaxy.

The distance is the dominant source of uncertainty in the calibration of the luminosity of SN~2008ax, as there is some discrepancy among published distances of NGC~4490. Here we adopt a distance of $9.6\pm 1.3$ Mpc ($\mu=29.92\pm0.29$ mag), derived by P08 averaging several different estimates, including Tully-Fisher, Sosies and kinematic distances.

\begin{figure*}
	\includegraphics[width=11.6cm]{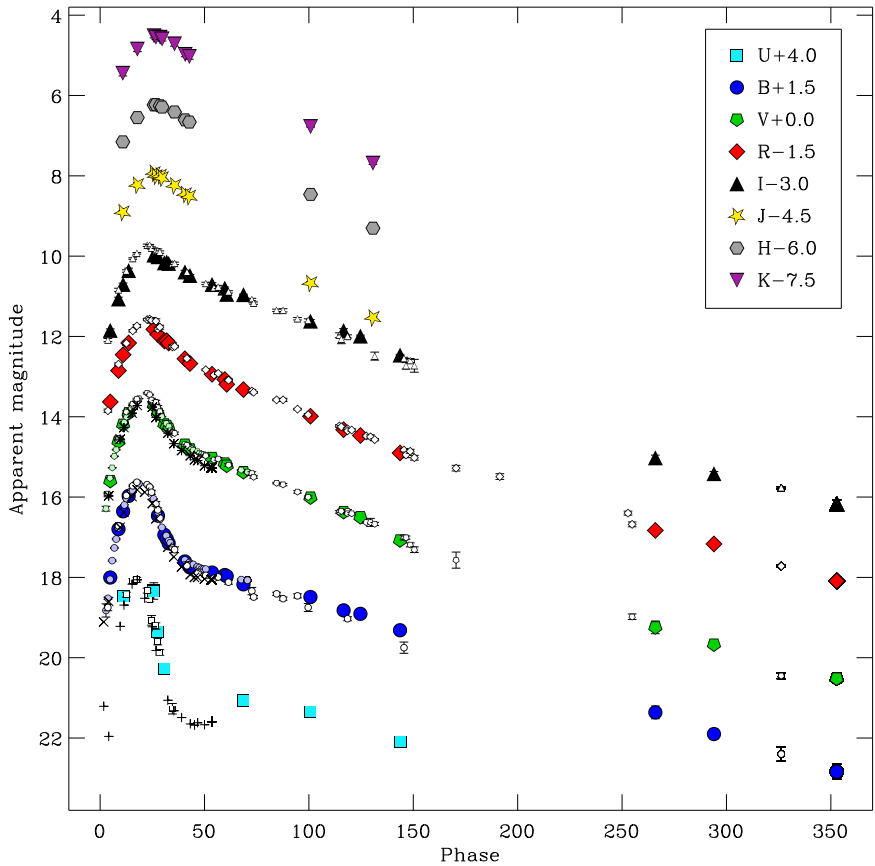}
	\caption{S-corrected $U\!BV\!RIJHK$ light curves of SN~2008ax. The phase is computed with respect to the explosion date (JD = $2\,454\,528.8 \pm 0.15$). $B$-band maximum light occurred on day 18.3. Small lightly coloured symbols are $B$- and $V$-band data from P08, small open symbols are $U\!BV\!RI$ data from Ts09, and pluses, crosses and asterisks are the $ubv$ photometry from R09.}
	\label{fig:sn08lc}
\end{figure*}

Another source of error is the unknown extinction along the line of sight towards SN~2008ax. Galactic reddening in the direction of NGC~4490 [$E(B-V) = 0.022$ mag, \citealt{sch98}] is very small compared to the reddening inside the host galaxy. Our measurement of the equivalent width (EW) of the Na I D lines ($1.8\pm0.1 $\,\AA), averaged over 20 epochs, is in good agreement with published numbers (P08, \citealt{Ch1}). Based on the EW of Na I D, \citet{Ch1} and P08 estimated colour excesses $E(B-V)= 0.5$ mag and $E(B-V)= 0.3$ mag, respectively. Ch10 showed that the Na I D lines are saturated, and derived a colour excess of $E(B-V)= 0.5$ mag using the EW of the K I $\lambda7699$ absorption. Here we adopt $E(B-V)=0.4 \pm 0.1$ mag as the total colour excess towards SN~2008ax, accounting for the facts that the empirical relation of \citet{t1}, used by P08, only gives a lower limit for the reddening, and that a closer match to the colour curves of SN~1993J is achieved with slightly larger reddening [adopting $E(B-V)= 0.2$ mag for SN~1993J; \citealt{m2}].

\section{Photometric analysis}

\subsection{Observed Light Curves}

The $U\!BV\!RIJHK$ magnitudes, acquired from five days to almost one year after the supernova explosion, are tabulated in Tables~\ref{opt_mags} and \ref{IR_mags}. The reported uncertainties are a quadratic combination of the measurement errors of the instrumental SN magnitudes and the errors associated with the photometric calibration. In Fig.~\ref{fig:sn08lc} we compare our light curves of SN~2008ax with Johnson $B$- and $V$-band data points from P08 (small lightly coloured symbols), $U\!BV\!RI$ data from Ts09 (small open symbols) and $ubv$ observations from R09 (pluses, crosses, asterisks). The P08 points are in very good agreement with our photometry, any systematic offset being smaller than the scatter of the light curves. Our $U$-band photometry is a bit brighter than that of Ts09, whereas in other bands our measurements are fainter, at least during the peak phase. The differences are time-dependent: large at early phases and smaller later on. The differences at peak are 0.10, 0.14, 0.19 and 0.19 mag in $B$, $V$, $R$ and $I$, respectively. A comparison of the magnitudes of the single sequence star we have in common with Ts09 shows that most of this difference (at least the time-independent component) comes from a deviant local-sequence calibration and hence different photometric zero points. The SWIFT $u$ band is bluer than most ground-based $U$ bands, and some differences between our and the R09 $U$-band photometry may be expected. The apparent re-brightening in the combined $U$-band light curve about 50\,d after the explosion may thus be an artefact caused by different passbands. The R09 $b$- and $v$-band light curves are slightly fainter than ours after maximum light, but agree within the measurement errors. In the end, most time-dependent differences can be related to different filter transmissions. Note that no $S$-correction has been applied to the photometry of P08, R09 and Ts09.

The epoch of the light-curve peak and the observed peak magnitude in each filter were estimated using moderate-order (3--5) polynomial fits. In Table~\ref{LC_param} the derived light-curve parameters are presented. SN~2008ax reached $V$-band maximum on UT March 23 at m$_{V} = 13.55$. Adopting a distance modulus of 29.92 mag, and a total extinction $A_V = 1.24$ mag, the absolute $V$ magnitude was $-17.61$ at peak.  This value can be considered normal for a member of this inhomogeneous class of objects \citep{r06}. It coincides with the second peak of SN~1993J (M$_{V} = -17.58$ for the distance and reddening adopted by \citealt{m2}) and the peak of the type IIb hypernova 2003bg (M$_{V} = -17.50$, \citealt{hamuy09,mazzali09}). At the same time, it is $\sim\,0.5$ mag and $\sim\,1$ mag brighter than the type Ib SNe~2008D \citep{maz08} and 2007Y \citep{s3}, respectively. About 60\,d after the explosion, SN~2008ax settles onto a linear decline with a rate of 1.90 mag per 100\,d in the $V$ band (see Table~\ref{LC_param}). This rate and those derived for other bands are similar, but somewhat faster, than the decline of SN~1993J. A least-squares fit to the data of SN~1993J from \citet{ba1} gives slopes for the $B$, $V$, $R$ and $I$ light curves of 1.46, 1.73, 1.57 and 1.77 mag per 100\,d, respectively, between 60 and 300\,d after the explosion. Except for rare cases (e.g. SN Ib 1984L; \citealt{sk89}) where the slope approaches the rate expected for $^{56}$Co decay in case of complete $\gamma$-ray trapping (0.98 mag per 100\,d), a steeper decline is common for SE-SNe between $\sim100$ and $300$\,d [e.g. SN Ib 2007Y \citep{s3}, SN Ib 1990I \citep{e1}, SN Ib 1983N, SN Ic 1983V, SN Ic 1994I \citep{cl1}, SN IIb 1993J \citep{richmond96,ba1}, and SN Ib 1996N \citep{sol98}], suggesting in general rather low ejecta masses compared to SNe~IIP.

\begin{table*}
	\centering
	\begin{minipage}{120mm}
	\caption{Light-curve parameters of SN~2008ax.}
	\label{LC_param}
	\begin{tabular}{@{}llcccc@{}}
	\hline
	Filter & Peak time\footnote{Based on a polynomial fit.} & Peak observed & Peak absolute & Decline rate of & $\Delta m_{15}$ \\
	&    (days after explosion) & magnitude & magnitude\footnote{Distance modulus $\mu = 29.92\pm0.29$ mag, colour excess $E(B-V) = 0.4\pm0.1$ mag.}&  radioactive tail\footnote{Average decline rate between 100 and 300\,d after the explosion for $BV\!RI$, and between 60 and 150\,d for $UJHK$ (in mag per 100\,d).} & (mag)\footnote{Decline within 15\,d from peak.} \\
		\hline
		
		$U$\footnote{$U$-band parameters were calculated including the data points of R09 and Ts09.} & 
                      $16.7\pm 0.5$ & $14.09\pm0.06$&$-17.73\pm0.56$& $1.74\pm0.30$&$2.84\pm0.06$ \\
		$B$ & $18.3\pm 0.5$ & $14.23\pm0.05$&$-17.32\pm0.50$& $1.74\pm0.11$&$1.48\pm0.05$ \\
		$V$ & $20.1\pm 0.4$ & $13.55\pm0.03$&$-17.61\pm0.43$& $1.90\pm0.09$&$0.91\pm0.03$ \\
		$R$ & $21.5\pm 0.4$ & $13.26\pm0.04$&$-17.69\pm0.39$& $1.64\pm0.08$&$0.60\pm0.04$ \\
		$I$ & $22.4\pm 0.7$ & $12.92\pm0.03$&$-17.75\pm0.35$& $2.03\pm0.15$&$0.39\pm0.03$ \\
		$J$ & $25.3\pm 0.2$ & $12.47\pm0.01$&$-17.80\pm0.30$& $2.89\pm0.18$&$0.48\pm0.01$ \\
		$H$ & $27.0\pm 0.4$ & $12.24\pm0.01$&$-17.90\pm0.30$& $2.82\pm0.14$&$0.40\pm0.01$ \\
		$K$ & $27.0\pm 0.1$ & $12.04\pm0.01$&$-18.02\pm0.29$& $3.05\pm0.14$&$0.47\pm0.01$ \\
	 \hline
	
		\end{tabular}
 \end{minipage}	
\end{table*}

\begin{figure}
	\includegraphics[width=8.4cm]{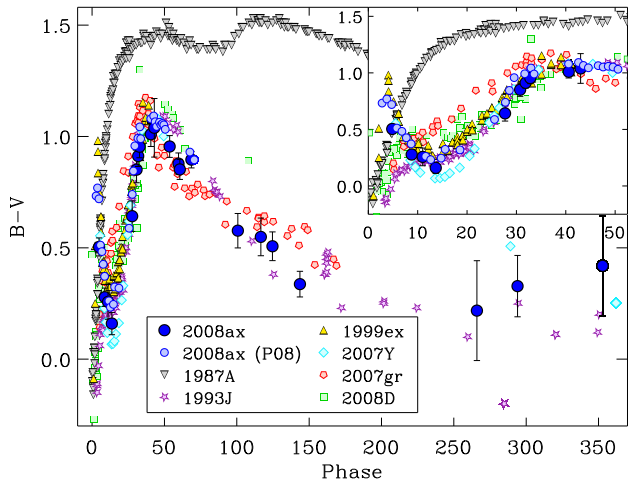}
	\caption{Comparison of the $B-V$ colour curve of SN~2008ax with those of other SNe. The phase is computed with respect to the explosion time. The colour excesses used for dereddening are shown in Table~\ref{SE-SNe}.}
	\label{fig:bv08ax}
\end{figure}

Fig.~\ref{fig:bv08ax} compares the $B-V$ colour evolution of SN~2008ax, the type Ic SN~2007gr \citep{h09}, the type Ib SNe~1999ex \citep{s2} and 2008D \citep{mod09}, the type IIb SNe~1993J \citep{vandriel93,lewis94,ba1,richmond96} and 2007Y \citep{s3}, and the type IIP SN~1987A (e.g. \citealt{cat89}; see also \citealt{wfh89} and references therein). The extinction towards SN~1999ex is rather uncertain. Here we adopt $E(B-V)=0.30$ mag \citep{s2}. The values for the other objects are the same as used for the bolometric curves (see Section~\ref{bolometric light curves} and Table~\ref{SE-SNe}). During the first two weeks after explosion, the $B-V$ colour curve of SN~2008ax closely resembles those of SNe~1999ex and 2007Y. These three SNe initially evolve towards bluer colours, suggesting a photospheric-temperature increase. Two weeks after the explosion, SN~2008ax reaches $B-V=0.1$ mag. Thereafter it turns redder, with a peak $B-V=1.1$ mag at $\sim$\,40\,d. At late phases the SN becomes bluer again, with $B-V=0.2$ mag on day 270. The close similarity of the colour curves of SNe~2008ax and 1993J from ten days after the explosion onwards is remarkable.
 
\subsection{Bolometric Light Curves}
\label{bolometric light curves}

\begin{figure}
	\includegraphics[width=8.4cm]{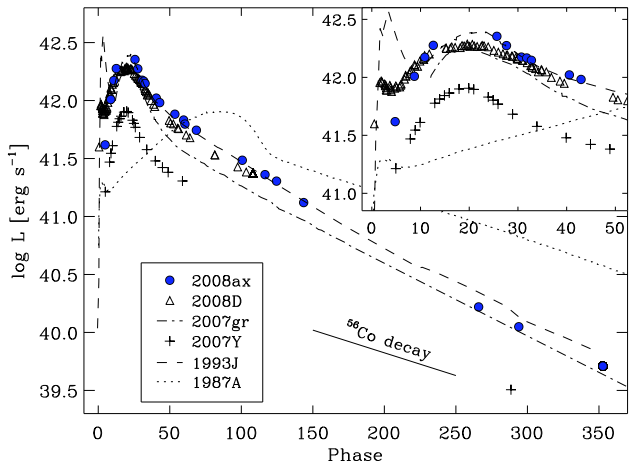}
	\caption{Comparison of the pseudo-bolometric light curves of SNe~2008ax (blue circles), 1993J (dashed line), 2008D (open triangles), 2007Y (crosses), 2007gr (dash-dotted line) and 1987A (dotted line). A reddening of $E(B-V)=0.4$ mag and a distance of 9.6 Mpc have been adopted for SN~2008ax. The phase is computed with respect to the explosion date.}
	\label{fig:sn08bol}
\end{figure}

We constructed the pseudo-bolometric light curve of SN~2008ax based on our optical $U\!BV\!RI$ and infrared $JHK$ data, assuming  $E(B-V)= 0.4\pm0.1$ mag and a distance of $9.6\pm1.3$ Mpc. To this aim, the magnitudes were first corrected for reddening and converted into monochromatic fluxes. The spectral energy distribution was then interpolated linearly and integrated over frequency. Finally, the integrated flux was converted into a luminosity using our adopted distance to NGC~4490. The extrapolation for missing $U$-band data at early and late phases, and missing $JHK$ coverage at late phases, was done assuming a constant colour with respect to the band closest in wavelength that covered these phases. 

Since SWIFT UV photometry exists for SN~2008ax (R09), we repeated the calculation of the pseudo-bolometric light curve including also these bands. The result, however, is quite uncertain owing to the extended red tail of the SWIFT $uvw1$ and $uvw2$ passbands, which for red objects shifts the effective wavelengths of the bands to the red. \citet{brown10} found that in type Ia SNe sometimes $\sim$\,90 per cent of the flux measured in $uvw1$ and $uvw2$ actually arises from the red tail. Lacking UV spectra of SN~2008ax, there is no possibility to determine proper red-tail corrections. The UV flux based on the magnitudes reported by R09 is thus only an upper limit to the true UV flux of SN~2008ax. Since even without a red-tail correction the UV contribution to the pseudo-bolometric light curve does not exceed $\sim$\,15 per cent at very early phases (when it is largest) and has dropped below 10 per cent by maximum light, we conclude that the UV does not give a major contribution to the luminosity of SN~2008ax at any phase, and neglect it hereafter.

In Fig.~\ref{fig:sn08bol} the pseudo-bolometric $U$-through-$K$-band light curve of SN~2008ax is compared to those of several well-studied SE-SNe and the type IIP SN~1987A. For SN~1987A the adopted distance modulus and total reddening are $\mu = 18.49$ mag and $E(B-V)=0.19$ mag, whereas for the SE-SNe the values are reported in Table~\ref{SE-SNe}. For SN~2008D we have used the numbers given by \citet{maz08}. SN~2007Y is assumed to have had a $B$-band rise time of 19\,d, similar to SNe~1999ex and 2008ax. The bolometric luminosity of SN~2008ax peaks at $\sim 2.4\times10^{42}$ erg\,s$^{-1}$ on day $\sim$20.7. Except for the first 10\,d, its light curve is very well matched by that of SN~1993J. The flux level at late times (after about 250\,d), however, is more similar to that of SN~2007gr.

\section{Spectroscopic analysis}

\subsection{Optical spectroscopy}
\label{Optical spectroscopy}

\begin{figure*}
	\includegraphics{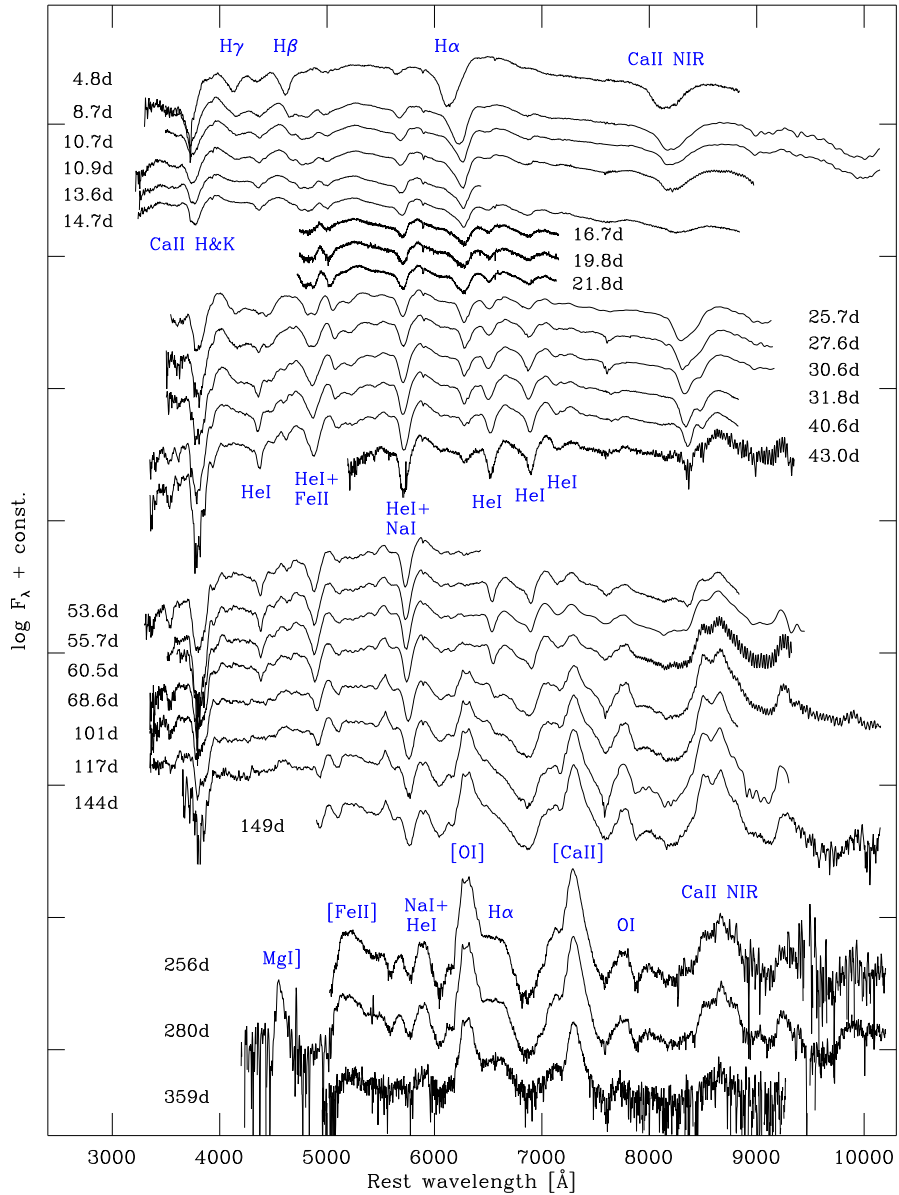}
	\caption{Spectroscopic evolution of SN~2008ax. The phase is computed with respect to the explosion date (JD\,$=2\,454\,528.8$). $B$-band maximum light occurred on day 18.3.}
	\label{fig:specev08}
\end{figure*}

The spectral sequence of SN~2008ax highlights that the sub-classification of SE-SNe is not straightforward, but may strongly depend on the phase at which a SN is discovered. 
The spectra of SN~2008ax show differences from those of the type IIb SN~1993J at early times. Similar to SN~1996cb \citep{q}, SN~2008ax was initially classified as a type II SN based on prominent Balmer P-Cygni features \citep{b1}. Only 5\,d after explosion it was re-classified as SN~IIb \citep{Ch1}, as the spectra showed increasing He I features and strong polarisation in H$\alpha$, more typical of type IIb than normal type II SNe at this early stage. By maximum light, He I lines dominated the optical and near-infrared spectra, which then closely resembled those of SNe~Ib such as SN~1999ex. 

\begin{figure}
\center
\includegraphics[width=6.8cm]{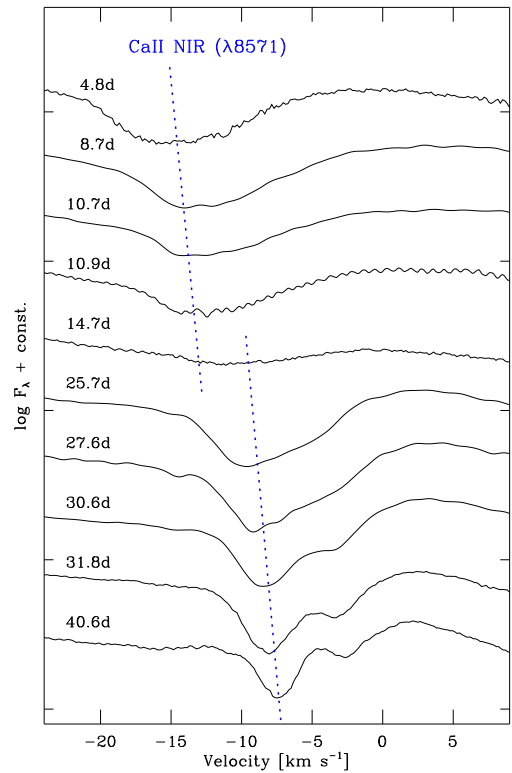}
\caption{Evolution of the Ca II NIR line of SN~2008ax in velocity space. Velocities are measured with respect to 8571 \AA. The flux is normalised to the local continuum, and a constant shift is applied.}
	\label{fig:Caii_NIR_ev}
\end{figure}

In Fig.~\ref{fig:specev08} we show our sequence of spectra, ranging from 5\,d to 1\,yr after the explosion. The earliest phases are dominated by P-Cygni features of the hydrogen Balmer series and Ca II. Evolving through maximum, the strength of H$\alpha$ steadily decreases, possibly overwhelmed by increasing He~I $\lambda6678$. At this phase, the SN was re-classified as type Ib \citep{tb1,ma1}. After maximum, the spectrum is dominated by strong He~I lines, with the feature near 5700\,\AA\ probably being a blend of He~I $\lambda5876$ and Na I $\lambda\lambda5890,5896$.  
The depth of the Ca~II NIR triplet absorption decreases until 15\,d after the explosion and increases again thereafter (Fig.~\ref{fig:Caii_NIR_ev}). This phenomenon was already observed in other SNe and interpreted as the transition from a high-velocity component to a photospheric component \citep[e.g.][]{s3,fol}. All in all, the photospheric phase lasts for about two months. 
Later, during the nebular phase, the spectra of SN~2008ax share similarities with those of SN~1993J, including prominent [O~I] $\lambda\lambda6300,6364$ emission and a boxy feature redwards of this line. The strong asymmetric emission peak at 4549\,\AA\ on day 280 can be identified as Mg~I] $\lambda4571$. 

\begin{figure*}
\includegraphics{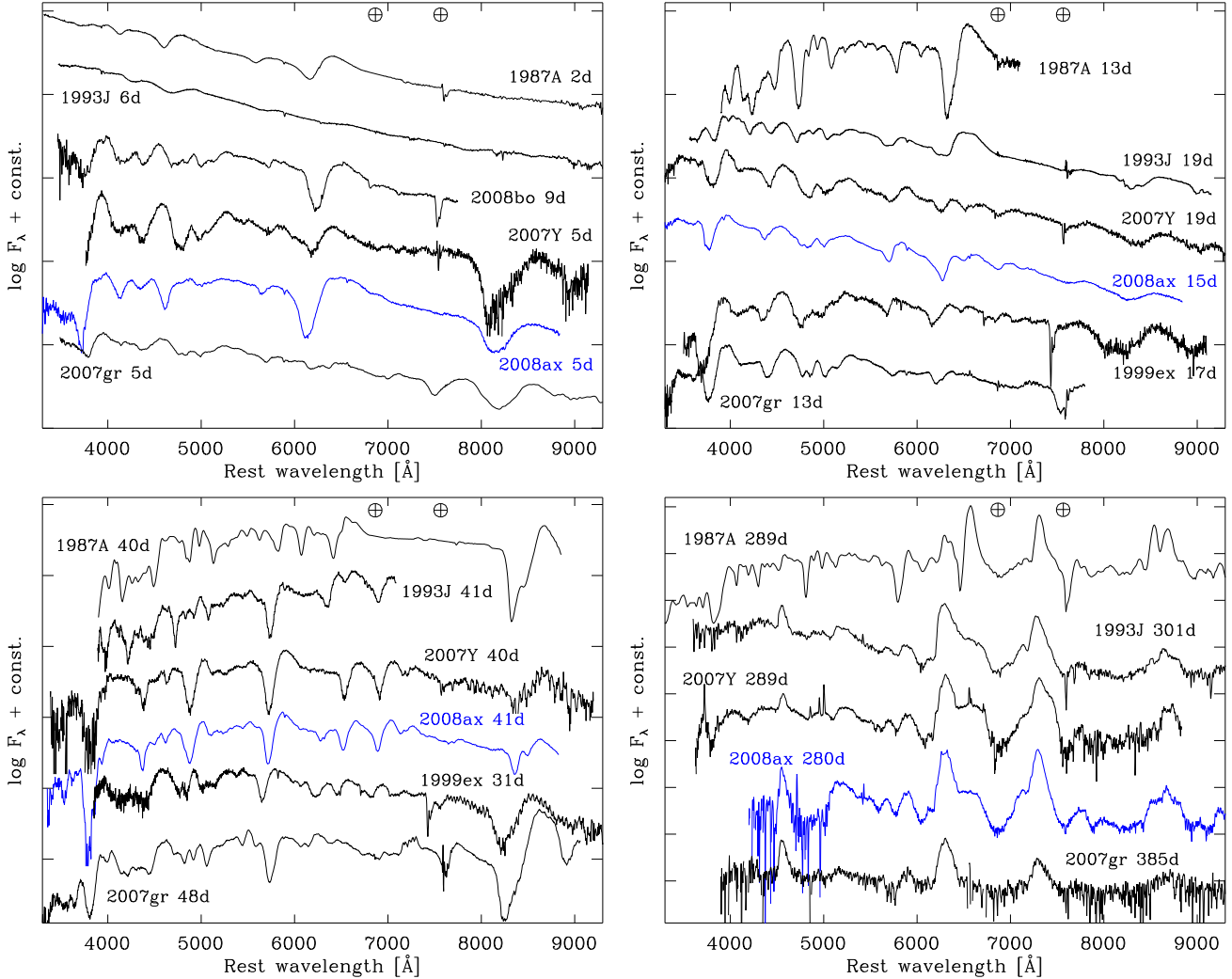}	
\caption{Comparison of SN~2008ax spectra with those of other core-collapse SNe at similar phases. The spectra are in the rest frame and have been dereddened assuming $E(B-V)=0.24$ mag for SN~2008bo, $E(B-V)=0.19$ mag for SN~1987A, and the values listed in Table~\ref{SE-SNe} for the remaining objects. The phase is computed with respect to the explosion time. $B$-band maximum occurred on day 18.3.}
\label{fig:08axcomp}
\end{figure*}

Fig.~\ref{fig:08axcomp} shows a comparison of the spectra of SN~2008ax at five days (upper-left panel), two weeks (upper-right), six weeks (lower-left) and $\sim$\,300 days (lower-right) after the explosion with those of the type IIP SN~1987A, 
the type IIb SNe~1993J \citep{ba1}, 2007Y \citep{s3} and 2008bo (Asiago SN archive), the type Ib SN~1999ex \citep{h02}, and the type Ic SN~2007gr \citep{v1,h09} at similar phases.  

At early epochs, SN~2008ax is similar to SNe~2007Y and 2008bo, but exhibits more prominent H$\alpha$. At 5\,d, SN~2008ax has an expansion velocity (deduced from the H$\alpha$ absorption minimum) of $\sim 20\,000$ km\,s$^{-1}$, compared to $\sim 16\,000$ km\,s$^{-1}$ in SN~2007Y. SN~1993J shows a blue, almost featureless continuum at this epoch, and in SN~2007gr no H$\alpha$ can be identified.  
Near maximum, SN~2008ax resembles type Ib/c rather than type II SNe, exhibiting higher expansion velocities and little hydrogen in its spectra. Helium lines are not yet fully developed at this phase. 
Six weeks after the explosion, the He I $\lambda\lambda4472,5876,6678,7065,7281$ lines are conspicuous, and SN~2008ax resembles a normal type Ib event with very weak H$\alpha$. At a similar phase, SN~2007Y shows no trace of H$\alpha$ near 6300\,\AA. 
 
With strong Mg~I] $\lambda4571$, [O~I] $\lambda\lambda6300,6364,$ [Ca~II] $\lambda\lambda7291,7324$ emission lines, blended [Fe~II] lines at $\sim5000$\,\AA\ and a characteristic, boxy feature on the red wing of the [O~I] doublet, the nebular spectra of SN~2008ax share similarities with those of SNe~1993J and 2007Y. For SN~1993J the boxy line profile has been explained by H$\alpha$ emission from a shell of hydrogen, possibly excited by interaction with a dense CSM \citep[e.g.][]{p95,hf}. A similar feature can also be seen in SNe~2007Y and 2008bo \citep{mi} during the nebular phase \citep{mi}, and thus seems to be common in SNe~IIb. The flux ratio [Ca~II]/[O~I] of SN~2008ax is slightly larger than in SN~1993J, and comparable to that of SN~2007Y.  Fransson \& Chevalier (1987, 1989) have shown that [Ca~II]/[O~I] is weakly dependent on the density and temperature of the emitting zone and remains relatively constant at late epochs. Besides, it seems to be sensitive to the core mass, thus tracing the main-sequence mass of the progenitor. The ratio increases with decreasing main-sequence mass. The [Ca~II]/[O~I] ratio of SN~2008ax ($\sim0.9$ at day 359) suggests a rather low-mass progenitor in a binary system rather than a single massive WR star, in agreement with the moderate mass-loss rate observed by R09.

\begin{figure}
\includegraphics[width=8.4cm]{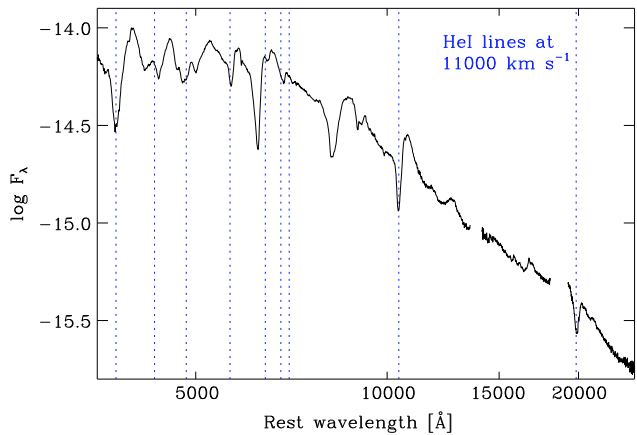}
\caption{Combined optical and NIR spectrum of SN~2008ax taken 11\,d after the explosion. Vertical dotted lines show the expected positions of He~I features for an expansion velocity of 11\,000 km\,s$^{-1}$.}
	\label{fig:08axoir}
\end{figure}

\subsection{NIR spectroscopy}

While the knowledge of the properties of SNe at optical wavelengths has made significant progress in recent years, still little is known in the infrared spectral range.  
Fig.~\ref{fig:08axoir} shows our earliest combined optical/NIR spectrum taken on 2008 March 14, about 11\,d after the explosion. Vertical lines mark the expected positions of He~I features for an expansion velocity of 11\,000 km\,s$^{-1}$. The spectrum shows strongly developed NIR He~I $\lambda\lambda1.083,2.058\mu$m lines, whereas optical He~I lines are still weak at this epoch.\footnote{This comes not unexpected, since the NIR He~I lines are transitions between (singlet or triplet) 2s and 2p levels. Being the lowest excited levels of He~I, especially the meta-stable 2s levels are strongly populated. The optical He~I transitions, on the other hand, all involve levels with $n \geq 3$.} The good agreement in the expansion velocities measured in all He~I lines and the very pronounced He~I $\lambda 2.058\mu$m feature suggest that the strong P-Cygni feature at $\sim1.05\mu$m is indeed dominated by He~I $\lambda1.083\mu$m, although some contribution of Paschen $\gamma$, C~I, Mg~II and other ions might be possible. Unfortunately, Pa$\alpha$ lies in a region where the earth's atmosphere is opaque. A weak feature at 1.24 $\mu$m, which was also observed in SN~1993J \citep{matt02}, is probably a hint of Pa$\beta$. As the suspected Pa$\beta$ is already weak, we do not expect a large contribution of Pa$\gamma$ to He~I $\lambda1.083\mu$m. 

\begin{figure*}
\center
\includegraphics{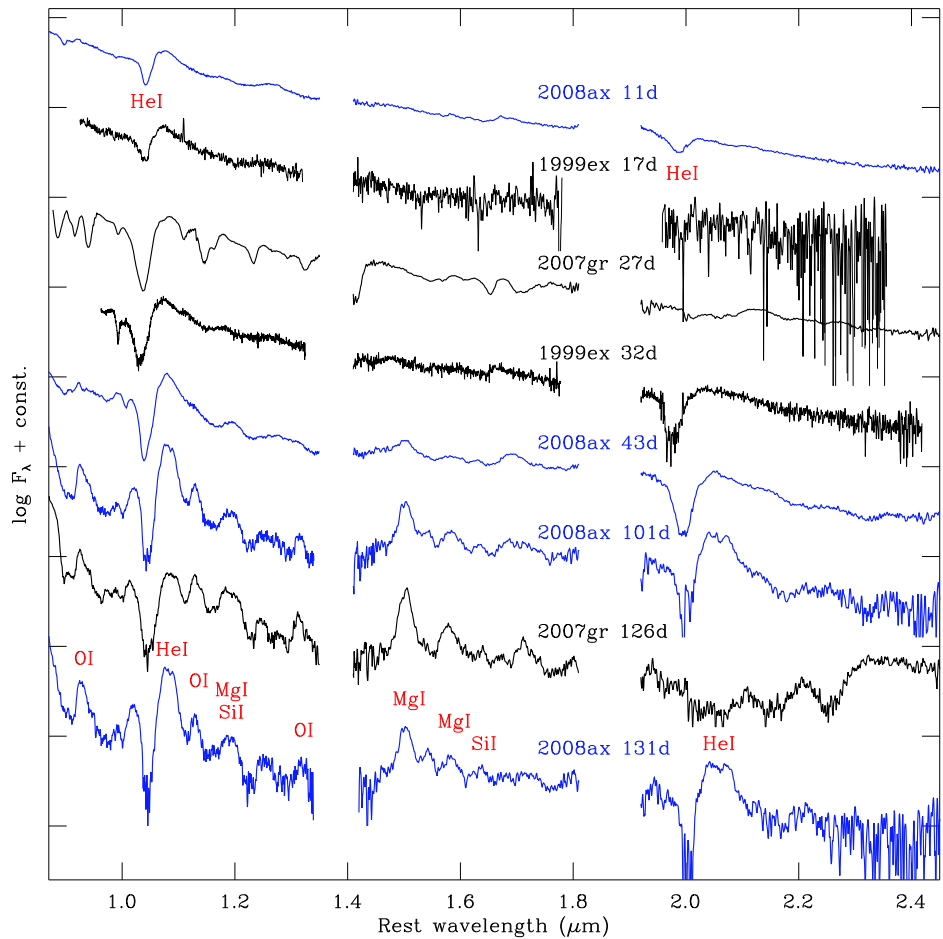}	
\caption{NIR spectra of SN~2008ax, compared with SNe~1999ex and 2007gr at epochs from 11 to 131 days after the explosion. The most prominent features have been labelled, following the line identifications given by \citet{maz10}.}
	\label{fig:08axinfcomp}
\end{figure*}

In Fig.~\ref{fig:08axinfcomp} we compare our four NIR spectra of SN~2008ax with spectra of the type Ib SN~1999ex \citep{h02} and the type Ic SN~2007gr \citep{h09} acquired at similar phases. All spectra are redshift-corrected and dereddened with the same values as adopted for the optical data. The most prominent features of the NIR spectra of SN~2008ax at all epochs are the strong He~I $\lambda\lambda1.083,2.058\mu$m lines. In that respect, SN~2008ax shows much closer resemblance to SN~1999ex than SN~2007gr, which lacks any hint of He~I $\lambda 2.058\mu$m as do all SNe~Ic \citep{tb06}. In quasi-nebular spectra, the emission bands at $\sim 0.93$ and $1.13\ \mu$m are attributed to O~I $\lambda0.926\mu$m and O~I $\lambda1.129\mu$m, respectively. Some contribution of [S~I] $\lambda1.131\mu$m to the band at $1.13\ \mu$m is likely \citep{maz10}. Three emission bands between 1.15 and 1.35 $\mu$m can be attributed to Mg~I $\lambda1.183\mu$m + Si~I $\lambda\lambda1.198,1.203\mu$m, [Fe~II] $\lambda\lambda1.257,1.279\mu$m and O~I $\lambda1.315\mu$m, respectively \citep{maz10}. Mg~I also contributes at 1.502, 1.575 and 1.711 $\mu$m, but the 1.7 $\mu$m feature is probably blended with He~I $\lambda1.700\mu$m, [Fe~II] $\lambda1.711\mu$m and [Co~II] $\lambda\lambda1.728,1.736\mu$m.
Contrary to SN~2007gr, no emission from CO molecular bands at $\lambda \geq 2.25\ \mu$m is observed in SN~2008ax at these epochs.

\subsection{Ejecta velocities}

Here we present a more detailed and quantitative study of the expansion velocities measured from relatively isolated lines.
All velocities have been determined measuring the blueshift of the minimum of the P-Cygni features, which is done by fitting a Gaussian profile to the absorption component in the redshift-corrected spectra. This procedure yields a rough estimate of the expansion velocities of the layers where the individual lines predominantly form. 

\begin{figure}
\includegraphics[width=8.4cm]{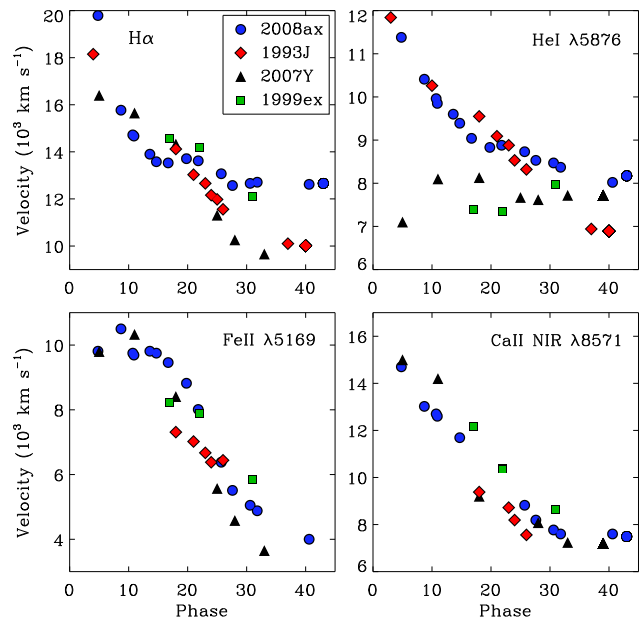}
\caption{Evolution of H$\alpha$, He I $\lambda$5876, Fe II $\lambda$5169 and Ca II NIR line velocities. The data of SN~1993J are from \citet{ba1} and \citet{o94}, those of SN~1999ex from \citet{h02}, and those of SN~2007Y from \citet{s3}. The H$\alpha$ identification in SN~1999ex is only tentative. The phase is computed with respect to the explosion time.}
	\label{fig:vel_comp}
\end{figure}

A comparison of the H$\alpha$, He~I $\lambda$5876,  Fe~II $\lambda$5169 and Ca~II NIR triplet line velocities of SN~2008ax with those of SN~1993J \citep{ba1,o94}, SN~1999ex (measured in spectra of \citealt{h02}) and SN~2007Y \citep{s3} is shown in Fig.~\ref{fig:vel_comp}. Ca~II NIR triplet velocities are measured with respect to 8571\,\AA\ and should be used with caution, as this feature apparently has two components. A high-velocity component at $\sim15\,000$ km\,s$^{-1}$ fades quickly and has disappeared by day 15, whereas a photospheric component at $\leq 9000$ km\,s$^{-1}$ develops just thereafter (Fig.~\ref{fig:Caii_NIR_ev}). This phenomenon has also been observed in SNe~2005bf \citep{fol} and 2007Y \citep{s3}.

SN~2008ax shows a large initial H$\alpha$ expansion velocity and a rapid decrease thereof during the first 15 days, with a drop from $\sim20\,000$ to $\sim12\,500$ km\,s$^{-1}$. The velocity then remains almost constant. The evolution of the expansion velocities measured from He~I $\lambda$5876 is similar to that of H$\alpha$, but with smaller initial velocities of $\sim12\,000$ km\,s$^{-1}$. After 20\,d it settles at $\sim8500$ km\,s$^{-1}$. The H$\alpha$ and He expansion velocities of SN~1993J exhibit a similar evolution as those of SN~2008ax before day 20, but continue to decrease until day 35. Instead, SN~2007Y shows an increase of the He velocity at early phases, followed by a slow decrease from $\sim$8500 to $\sim$7500 km\,s$^{-1}$. This non-uniform behaviour of He~I $\lambda$5876 velocities in SE-SNe may be caused by variable degrees of contamination with Na~I $\lambda\lambda5890,5896$. 
The evolution of the expansion velocities of Fe~II $\lambda$5169, which is often used as a tracer of the photospheric velocity, is in good agreement among the four SE-SNe considered here. In early spectra, Fe~II velocities of $\sim10\,000$ km\,s$^{-1}$ are measured, decreasing to $\sim4000$ km\,s$^{-1}$ six weeks after the explosion.

\section{Discussion}

\subsection{Explosion parameters}

\subsubsection{Light-curve modelling}
\label{Light-curve modelling}

\begin{figure}
	\includegraphics[width=8.4cm]{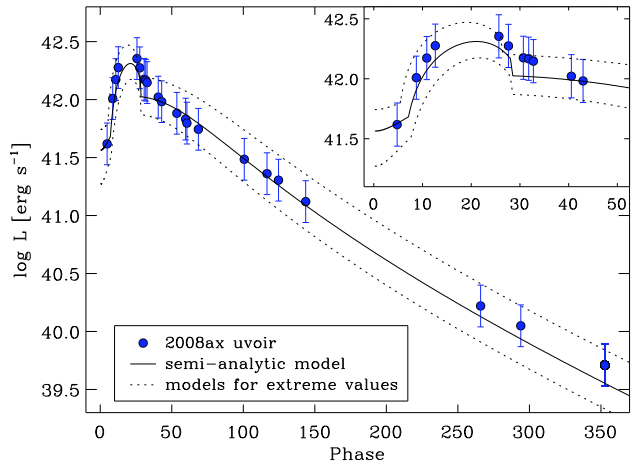}
	\caption{Observed (points) and computed (solid line) bolometric light curves. Dotted lines are model light curves built for extreme distance and reddening values. The phase is computed with respect to the explosion date. The large error bars are due to distance and reddening uncertainties.}
	\label{fig:08axbolfit}
\end{figure}

The physical properties of the envelope have been derived using a semi-analytic code (see \citealt{zam03} and \citealt{zam07} for details), which performs a simultaneous $\chi^{2}$ fit of three observable quantities (the bolometric light curve, the evolution of the photospheric velocity and the continuum temperature at the photosphere) with model calculations. The code assumes a homologously expanding SN envelope of uniform density and spherical symmetry. With the explosion epoch, distance modulus and reddening adopted in this paper, a good fit is obtained with an initial radius $R_{0} \simeq 3 \times 10^{11}$ cm ($\sim4\ R_\odot$), a velocity of the expanding envelope of about 10\,000 km\,s$^{-1}$, an explosion energy of $E_{0} \sim 6\times10^{51}$ erg, a $^{56}$Ni mass of $\sim0.1\ M_\odot$ and a total ejected mass of $M_\mathrm{ej} \simeq 4.5\ M_\odot$. The computed bolometric light curve is shown in Fig.~\ref{fig:08axbolfit} (solid line). It fits the radioactive tail quite well, but slightly underestimates the maximum. Large uncertainties related to distance and reddening cause uncertainties in the estimated physical parameters. These have been determined calculating models for extreme assumptions on distance and reddening: small distance with low reddening (lower dashed line) and large distance with high reddening (upper dashed line). Moreover, being a one-zone model our light-curve fit tends to overestimate the ejecta mass (and hence the kinetic energy) to compensate for the reduced opacity.  As a result of all these considerations, we estimate the ejecta mass to be between $2$ and $5\ M_\odot$, the $^{56}$Ni mass to be in the range 0.07--0.15 $M_\odot$, and the explosion energy to be between 1 and 6 foe (1 foe is $10^{51}$ erg).

\subsubsection{Modelling of the early-time H$\alpha$ absorption}

\begin{figure}
\includegraphics[width=8.4cm]{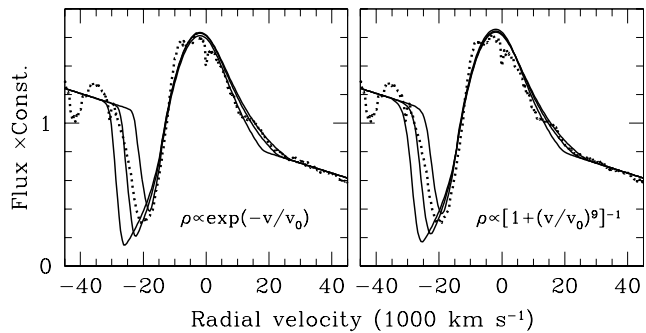}
\caption{ Comparison of the observed H$\alpha$ profile (dotted line) in the 4.8\,d spectrum with the models described in the text. Models with an exponential density profile are shown on the left, those with a power-law density profile on the right. In each panel the solid lines represent models with 1.0, 1.5 and 2.0 foe. Higher H$\alpha$ velocities correspond to larger total kinetic energy of the ejecta. }
	\label{fig:Halpha_early}
\end{figure}

The high-velocity H$\alpha$ absorption wing 4.8\,d after the explosion is sensitive to 
the hydrogen density in the outermost layers of the ejecta, which in turn depends on the
mass and kinetic energy. One can use this fact to constrain independently these ejecta 
parameters. We use two versions of the density distribution $\rho(v)$, an exponential
profile $\rho\propto\exp(-v/v_0)$ and a plateau with  an external power-law decay 
$\rho\propto[1+(v/v_0)^9]^{-1}$. Both distributions seem reasonable: the exponential 
one approximates supernovae arising from compact progenitors, while the power law is 
more adequate in case of SNe~IIP with extended hydrogen envelopes, e.g. SN~1987A 
\citep{Utrobin07}. The population of the $n=2$ level of hydrogen in the model is 
determined by recombination, Ly$\alpha$ scattering and two-photon decay. This is 
a reasonable approximation for the outer layers. The hydrogen abundance is set to be 
solar, while the ionisation fraction ($x=0.81$) is chosen such that the $n=2$ level 
population is maximised. This choice minimises the hydrogen mass in the outer layers 
required in our model to produce the high-velocity H$\alpha$ absorption wing. We 
fixed the ejecta mass at $4~M_{\odot}$ (as a compromise between the results by R09, 
Ts09 and the light-curve modelling in this work) and varied only the kinetic energy 
between 1 and 2 foe (see Fig.~\ref{fig:Halpha_early}). 

For both the exponential and power-law density distribution the blue absorption wing 
suggests that $E_\mathrm{kin}=1.5$ foe is a lower limit for the total kinetic energy of 
the ejecta. If the hydrogen fraction was lower or the degree of ionisation different, 
a larger kinetic energy would be required to reproduce the highest velocities in 
H$\alpha$. The adopted mass and the derived minimum energy are consistent with the 
parameters estimated by Ts09 and M10, as well as the light-curve modelling in 
Section~\ref{Light-curve modelling}. The low energy of 0.5 foe found by R09, however, 
is not supported by our modelling of the early H$\alpha$ absorption.\\

\begin{figure}
\includegraphics[width=8.4cm]{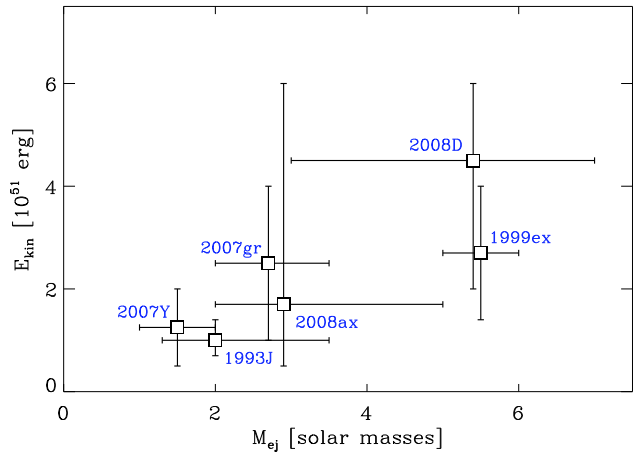}
\caption{Kinetic energies of stripped-envelope CC-SNe as a function of their ejecta masses. If no errors are given in Table~\ref{SE-SNe}, we have assumed uncertainties of $\pm30$ per cent on the ejecta mass and $\pm50$ per cent on the total kinetic energy. A weak correlation between these two quantities can be discerned.}
\label{fig:E_vs_M}
\end{figure}

In Table~\ref{SE-SNe} we show the physical parameters of SN~2008ax and other SE-SNe, 
derived using a variety of methods. SN~2008ax appears to be similar to SNe~1993J and 
2007Y in terms of $^{56}$Ni and ejecta mass. A plot of the kinetic energy as a function 
of ejected mass (Fig.~\ref{fig:E_vs_M}) shows the expected correlation between these two 
quantities within the (admittedly large) uncertainties. It also suggests that type IIb 
SNe generally have ejecta masses between 1 and 5 $M_{\odot}$ and explosion energies 
between 0.5 and 6 foe.

\begin{table*}
\centering
	\begin{minipage}{176mm}
	\caption{Properties of various SE-SNe.}
	\label{SE-SNe}
	\begin{tabular}{@{}lcccccccc@{}}
	\hline
    SN & Type   &M$_{B,\mathrm{max}}$ & $\mu$\footnote{Distance to SN~2008ax from P08, obtained averaging the results from several methods; short distance to SN~2008D ($\mu=32.16$ mag) from the Tully-Fisher relation; distances to SNe~1993J and 2007gr from Cepheids; kinematic distances used for SNe~1999ex, 2007Y and 2008D (long distance, $\mu=32.45$ mag).}         & $E(B-V)_\mathrm{tot}$      & $^{56}$Ni mass & Ejecta mass & $E_\mathrm{kin}$ & Reference \\
       &        &                 & (mag)          & (mag)         & ($M_\odot$)    & ($M_\odot$)  & (10$^{51}$ erg)  &           \\
\hline
2008ax & IIb    & $-17.32\pm0.50$ & $29.92\pm0.29$ & $0.4\pm0.1$   & 0.07--0.15    & 2--5        & 1--6           & This work\\	
       &        & $-17.32\pm0.50$ & $29.92\pm0.29$ & $0.4\pm0.1$   & 0.07--0.15    & 1.9--4.0    & 0.7--2.1       & \citet{M10}\\  
       &        & $-16.87$        & $29.92\pm0.29$ & 0.3           & 0.06          & 2.9         & 0.5            & \citet{r09}\\	
       &        & $-17.06$        & $29.92\pm0.29$ & 0.3           & 0.11          & 2.3         & 1.5            & \citet{ts09}\\
2008D  & Ib     & $-16.30$        & 32.16          & $0.6\pm0.2$   & 0.05--0.10    & 3--5        & 2--4           & \citet{sod08}\\
       &        &                 & 32.45          & 0.65          & 0.09          & 7           & 6              & \citet{maz08}\\
2007gr & Ic     & $-16.75$        & $29.84\pm0.16$ & $0.09\pm0.02$ & 0.06--0.10    & 2.0--3.5    & 1--4           & \citet{h09}\\
2007Y  & Ib/IIb & $-16.20$        & $31.43\pm0.55$ & 0.11          & 0.06          & 1--2        & 0.5--2.0       & \citet{s3}\\
1999ex & Ib/c   & $-17.42$        & $33.54\pm0.23$ & $0.30\pm0.04$ & 0.16          & 5--6        & 2.7            & \citet{s2}\\
1993J  & IIb    & $-17.23$        & $27.80\pm0.08$ & 0.2           & 0.10--0.14    & 1.9--3.5    & 1.0--1.4       & \citet{y95}\\
       &        &                 & $27.80\pm0.08$ & 0.2           & 0.10          & 1.3         & 0.7            & \citet{r06}\\ 
\hline
	\end{tabular}
 \end{minipage}	
\end{table*}

\subsection{Nature of the progenitor star}

Two different scenarios  for the progenitor of SN~2008ax have been proposed by \citet{c1} analysing pre-explosion HST observations: the precursor may have been a single massive star, which had lost most of its H-rich envelope and exploded as a helium-rich Wolf-Rayet star, or a stripped star in an interacting binary system. P08 favoured the explosion of a WNL star on the basis of the early spectro-photometric evolution of the SN. R09 instead suggested a binary progenitor based on UV, optical, X-ray and radio properties of SN 2008ax. 

With our estimated physical parameters, SN~2008ax lies between SN~1993J, whose progenitor is thought to be a relatively-low-mass star in a binary system \citep{Podsi93,y95,Maund09}, and SN~2008D, which probably had a massive Wolf-Rayet progenitor \citep{maz08,sod08,mod09,tan09}. However, the [Ca~II]/[O~I] ratio in nebular spectra of SN~2008ax (Section~\ref{Optical spectroscopy}) lends support to the lower-mass binary scenario.

\subsection{Late-time H$\alpha$ emission}
\label{Late-time Ha emission}

\citet{p95} found a late-time flux excess in the H$\alpha$ line of SN~1993J compared to radioactive models. They argued that the nebular H$\alpha$ luminosity was powered by interaction with a stellar wind from the progenitor. With a model for the ejecta-wind interaction, they derived a mass-loss rate of $\dot{M} = 2\times10^{-5}(v_\mathrm{w}/10$ km\,s$^{-1}$) $M_{\odot}$\,yr$^{-1}$ for the progenitor of SN~1993J. However, \citet{hf} argued that H$\alpha$ should be optically thick even during the nebular phase. Considering all contributions, in particular the scattering of photons emitted in the strong [O~I] $\lambda\lambda6300,6364$ feature by H$\alpha$, they did not detect a significant flux excess until $\sim$\,200\,d, suggesting no dominant contribution from ejecta-wind interaction up to that phase.
Similarly, \citet{ChS09} tried to explain the H$\alpha$ emission in nebular spectra of the SN~2007Y with radioactivity alone, claiming that circumstellar interaction should be undetectable at late phases because of the low density and large shock radius.

In SN~2008ax, observations of the H$\alpha$ luminosity evolution may hint at some contribution from shock interaction at the latest phases.
The H$\alpha$ luminosity evolution of SN~2008ax was determined as follows. At early epochs, the local continuum around H$\alpha$ was removed from the reddening-corrected spectra by subtracting a linear fit. The residual flux in the H$\alpha$ P-Cygni profile was then integrated over wavelength, and transformed into a luminosity adopting a distance of 9.6 Mpc. At nebular epochs, the emission immediately redwards of [O~I] $\lambda\lambda6300,6364$ was assumed to be H$\alpha$. Since the blue wing was contaminated by [O~I], the H$\alpha$ emission profile was assumed to be symmetric about the rest wavelength. The total flux is thus twice the value obtained by integration redwards of 6563\,\AA\ only.

\begin{figure}
\includegraphics[width=8.4cm]{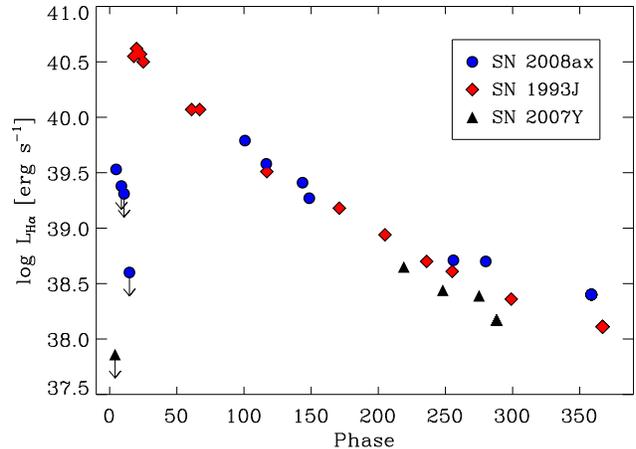}
\caption{ Comparison of the H$\alpha$ luminosity curve of SN~2008ax (blue circles; arrows denote upper limits) with those of SN~1993J from \citet{p95} (red squares) and SN~2007Y (black triangles). The phase is computed with respect to the explosion time. }
	\label{fig:Ha_comp}
\end{figure}

Fig.~\ref{fig:Ha_comp} shows a comparison of the H$\alpha$ luminosity curve of SN~2008ax with those of SNe~1993J \citep{p95} and 2007Y (derived from spectra published by \citealt{s3}). Adopted distances are reported in Table~\ref{SE-SNe}. Compared to SN~1993J, SNe~2008ax and 2007Y show almost no net emission in H$\alpha$ at early times. Instead, their H$\alpha$ lines have perfect P-Cygni scattering profiles. 
After about 100 days after the explosion, a broad feature arises redwards of [O~I] $\lambda\lambda6300,6364$, which can be tentatively identified with H$\alpha$. However, at least during the early nebular phase until about 150 days after the explosion, a contribution from various other elements, most notably Fe, is likely \citep{p95}. Moreover, at least part of the flux could arise from H$\alpha$ scattering rather than pure emission (\citealt{hf}, M10). 
Finally, there is a flattening in the H$\alpha$ luminosity curve at very late phases, after $\sim$\,300\,d for SN~1993J and $\sim$\,250\,d for SN~2008ax. In SN~1993J this was interpreted as the transition to an interaction-dominated phase. The similarity of the H$\alpha$ luminosity evolution of SNe~1993J and 2008ax suggests that the same mechanism is responsible for the emission in both objects.

The shock-wave mechanism for the late H$\alpha$ emission in SN~2008ax, however,
faces a serious problem:
the high velocity of the interface between the ejecta and wind at this stage. We
calculated the interaction dynamics in the thin-shell approximation, adopting
$E_\mathrm{kin}=1.5$ foe, $M=4~M_{\odot}$ for the SN ejecta, and a wind density  
$w\leq\dot{M}/v_\mathrm{w}=6.3\times10^{14}$ g\,cm$^{-1}$, which corresponds to 
a mass-loss rate of $\dot{M} \leq 10^{-5}~M_{\odot}$ yr$^{-1}$ for a wind velocity 
$v_\mathrm{w}=10$ km\,s$^{-1}$ (as estimated by Ch10).
If the H$\alpha$ line is emitted by the cool dense shell at the reverse shock, this 
model results in a boxy line profile with a characteristic radial velocity of $\geq 12\,000$ 
km\,s$^{-1}$ on day 359, at least twice as large as the velocity of $\sim$\,6000 
km\,s$^{-1}$ measured at the edge of the observed feature.

The mechanism of the H$\alpha$ emission powered by the shock wave could be salvaged 
if the hydrogen-rich ejecta or the surrounding circumstellar material were distributed 
in a disk, observed at a small inclination angle. The required angle is determined by the 
ratio of the observed line width and the model velocity of the thin shell: $i\sim30^{\circ}$. 
In this scenario, however, significantly broader H$\alpha$ emission (with edge velocities 
$> 10\,000$ km\,s$^{-1}$) is expected to be found in future observations of other SNe~IIb.

Alternatively, most of the emission is actually related to low-velocity hydrogen of the 
unshocked ejecta, located far from the shock. This scenario requires a hydrogen distribution 
with an inner cut-off at 5000 km\,s$^{-1}$, and a maximum hydrogen density attained at about 
this velocity. The mechanism of hydrogen emission in that case could include excitation by hard 
($\sim$\,$100$ keV) X-rays from the forward shock. However, owing to the low luminosity of 
the shock ($<10^{39}$ erg s$^{-1}$) and the small dilution factor ($\sim$\,0.06), the excitation 
by hard X-rays can provide only a minor contribution to the non-thermal excitation 
owing to $^{56}$Co decay with a $\gamma$-ray luminosity of $\sim$\,$3\times10^{40}$ erg s$^{-1}$.
Ly$\alpha$ scattering, however, might help by increasing the population of the second level, 
from which hydrogen can be ionised by SN radiation with $h\nu>3.4$ eV. The emission rate of 
H$\alpha$ quanta is then equal to the number of quanta absorbed from the Balmer continuum. 
This is another version of the ordinary non-thermal radiative mechanism of hydrogen emission 
in SNe~IIP.

\subsection{Explosion geometry}

\subsubsection{Nebular [O~I] line profile}

\begin{figure}
\includegraphics[width=8.4cm]{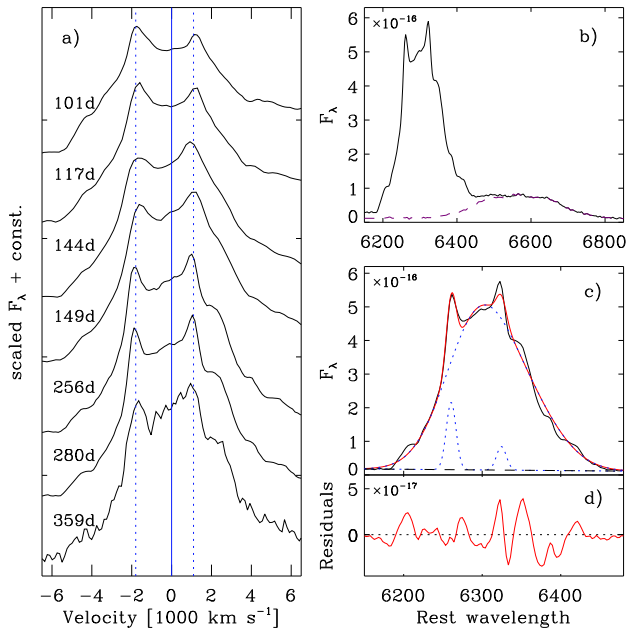}
\caption{a) Enlarged line profiles of [O~I] $\lambda\lambda6300,6364$ in 7 nebular spectra of SN~2008ax, shown in velocity space. The solid vertical line marks zero velocity with respect to 6300\,\AA. The dashed vertical lines mark the peaks blueshifted by about 1800 km\,s$^{-1}$ and redshifted by about 1200 km\,s$^{-1}$. b) [O~I] at 280\,d and the H$\alpha$ profile created assuming symmetry with respect to  6563\,\AA. c and d) 2-component Gaussian fit \citep[c.f.][]{tb09} and residuals thereof.}
\label{fig:OIprofile}
\end{figure}

What distinguishes SN~2008ax from SNe~1993J and 2007Y during the nebular phase is the symmetric, double-peaked profile of [O~I] $\lambda\lambda6300,6364$. Double-peaked emission lines in late-time spectra of SE-SNe are relatively frequent \citep{maeda08,mod08}, and a possible explanation is that oxygen has a torus-like distribution and our line of sight is near the plane of the torus \citep{maz05,maeda08,ta09,tb09,Maurer10}. A different model has been discussed by \citet{mi}, suggesting the double peak to be caused by the doublet nature of [O~I] $\lambda\lambda6300,6364$, with a line ratio close to 1. The atomic physics of the [O~I] doublet has been discussed by \citet{spypi91} and \citet{li92}, who show that in a supernova the ratio of [O~I] $\lambda6300$ to [O~I] $\lambda6364$ should evolve from $\sim1$ at early times to $\sim3$ at late times, as the ejecta expand and the lines become optically thin. This was actually observed in SNe~1986J \citep{lei91}, 1987A \citep{spypi91,li92} and 1988A \citep{spy91}. \citet{tb09} argue that for SE-SNe at 100\,d after the explosion the ratio should already be 3:1. Fig.~\ref{fig:OIprofile}a shows the evolution of the [O~I] line of SN~2008ax between 101 and 359\,d since explosion. One would expect to see a temporal intensity decrease of the red peak with respect to the blue one if the two peaks were due to optically thick [O~I] $\lambda6300$ and [O~I] $\lambda6364$, as seen in SN~1987A \citep{li92}. However, this ratio \textit{in}creases with time, as seen also in SN~2004ao \citep{mod08}. 

To exclude a contribution of H$\alpha$ flux to the redshifted peak of [O~I], we have subtracted the boxy profile assuming symmetry with respect to the H$\alpha$ rest wavelength (see Fig.~\ref{fig:OIprofile}b, dashed line). We have then fitted the H$\alpha$-free [O~I] profiles at four epochs (149--359\,d) adopting the multi-Gaussian fitting procedure described by \citet{tb09}. One-component and classical double-peak fits have shortcomings in that they do not reproduce the two narrow peaks. A satisfactory fit can be obtained using 2 components: a broad one approximately at the rest wavelength, with a blueshifted, narrow one superimposed. The FWHM of the broader component is about 4800 km\,s$^{-1}$, and it is systematically blueshifted by about 300 km\,s$^{-1}$. The narrow component with FWHM of about 700 km\,s$^{-1}$ is blueshifted by about 1900 km\,s$^{-1}$.  The relative flux of the second component ($\alpha_{2}$ in the notation of \citealt{tb09}) is about 0.07. The best 2-component fit for the 280\,d spectrum is presented in Fig.~\ref{fig:OIprofile}c, and the residuals are shown in Fig.~\ref{fig:OIprofile}d. The residuals might suggest an additional redshifted component, but the significance is rather low.
Taken at face value, the derived fit configuration indicates an almost central, spherically symmetric distribution of the bulk of the oxygen-rich ejecta with an expansion velocity of $\sim 4800$ km\,s$^{-1}$, and a clump with enhanced density and/or excitation at a line-of-sight velocity of $\sim -1900$ km\,s$^{-1}$. However, a thin torus viewed from an equatorial direction in addition to a spherically symmetric mass of oxygen, or an aspherical distribution of the $^{56}$Ni exciting the oxygen, might be possible alternatives. Finally, M10 suggested that the minimum between the two peaks of [O~I] may not have a geometric origin, but instead be produced by scattering in optically thick H$\alpha$, with the corresponding emission contributing to the late-time H$\alpha$ feature discussed in Section~\ref{Late-time Ha emission}.

\subsubsection{Nebular line profiles in the NIR}

\begin{figure}
\includegraphics[width=8.4cm]{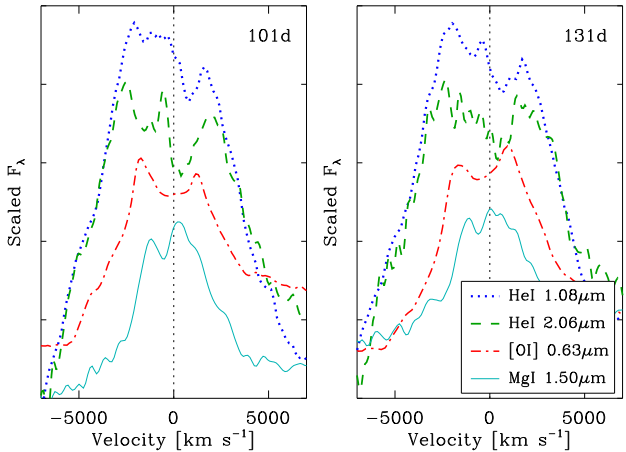}
	\caption{[O~I] $\lambda\lambda6300,6364$, He~I $\lambda1.083\mu$m, He~I $\lambda2.058\mu$m and Mg~I $\lambda1.502\mu$m profiles of SN 2008ax at 101 and 131 days after explosion. [O I] is plotted with respect to 6300\,\AA. As there is no optical spectrum on day 131, the [O~I] profile in the right panel is from a spectrum taken on day 144.}
	\label{fig:oiheinf_08ax}
\end{figure}

During the nebular phase, also the NIR He~I $\lambda\lambda 1.083,2.058\mu$m lines of SN~2008ax show double-peaked emission (Fig.~\ref{fig:oiheinf_08ax}). Actually, there are additional narrow peaks on day 101 in He~I $\lambda2.058\mu$m, blueshifted by about 450 km\,s$^{-1}$ with respect to the rest wavelength, and on day 131 in He~I $\lambda1.083 \mu$m, blueshifted by about 400 km\,s$^{-1}$. However, as these narrow peaks do not persist at both epochs, they might be reduction artefacts and we do not consider them hereafter. The good match of the two NIR He~I lines suggests that the observed double peak is indeed due to He~I and not caused by a blend with other lines. In Fig.~\ref{fig:oiheinf_08ax} we overlap the NIR He~I emission lines with Mg~I $\lambda1.502\mu$m and the double-peaked optical [O~I] $\lambda\lambda6300,6364$. Vertical lines mark zero velocity. The fluxes are scaled arbitrarily. The left panel shows the profiles at 101\,d, the right one at 131\,d ([O~I] $\lambda\lambda6300,6364$ at 144\,d). The expansion velocities of He~I are higher than those of O~I. We have measured a FWHM of $\sim7600$ km\,s$^{-1}$ for the He~I lines and a FWHM of $\sim5000$ km\,s$^{-1}$ for [O~I] (corrected for the doublet nature). The peaks of the He~I lines are centred at $-2000$ and $+1700$ km\,s$^{-1}$ with respect to the rest wavelength on day 131, while those of [O~I] are found at $-1700$ and $+1000$ km\,s$^{-1}$ with respect to 6300\,\AA\ on day 144. The Mg~I $\lambda1.502\mu$m line has a smaller FWHM ($\sim4000$ km\,s$^{-1}$), in good agreement with the value from Mg~I] $\lambda4571$ on day 280. The blueshifted peak is centred at $-1200$ km\,s$^{-1}$, the redshifted one close to the rest wavelength.\footnote{ The structure seen in Mg~I may, however, also be due to a blend of several Mg I lines ($\lambda\lambda1.488,1.504,1.505\mu$m).} A possible common explanation for all these double-peaked profiles might be non-uniform distribution of $^{56}$Ni within the ejecta, which induces different degrees of heating and ionisation/excitation in the surrounding material. The variation in the double-peak separation for different species would then be a consequence of the chemical stratification of the ejecta.

\section{Conclusions}

SN~2008ax is one of the rare cases of a SN~IIb discovered soon after the explosion and densely monitored from X-rays to radio waves.  
Our optical and NIR photometric and spectroscopic observations of SN~2008ax span almost one year from explosion. We have compared our data with previously published observations of SN~2008ax and with a sample of relatively well-studied SE-SNe. SN~2008ax appears to be a normal SE-SN, exhibiting properties in common with other SNe~IIb and Ib. Its light curves most closely resemble those of SN~1993J from day 10 onwards, reaching an absolute peak magnitude of M$_{V}=-17.6$. Similar to SN~1993J, the late decline rates are substantially faster than the decay rate of $^{56}$Co, indicating a leakage of $\gamma$-rays.  

The spectral similarities with SNe~2007Y and 1999ex at early phases suggest similar photospheric conditions. However, some differences are also observed, which are likely related to deviations in ejected mass and geometry. We argue that the double peaks in the [O~I] $\lambda\lambda6300,6364$ lines observed in spectra since $\sim$\,$100$\,d after the explosion have a geometric origin and are not caused by a $\sim$1:1 flux ratio of the doublet lines at high optical depth. This is consistent with significant late-time continuum polarisation found by Ch10. 
  
NIR spectra show strong He I lines similar to SN~1999ex. Nebular NIR spectra exhibit a large number of emission features: He~I, O~I, Mg~I, Si~I, [Fe~II]. Particularly interesting are the prominent, double-peaked profiles of the He~I $\lambda\lambda1.083,2.058\mu$m lines, which -- together with the profiles of [O I] $\lambda\lambda6300,6364$ -- provide a clue for an asymmetric large-scale Ni mixing in the ejecta. 

From modelling the bolometric light curve with a semi-analytic code, we have derived a total ejected mass of 4.5 $M_\odot$, 0.1 $M_\odot$ of which is $^{56}$Ni. Our model yields a progenitor radius $R_{0} \simeq 3 \times 10^{11}$ cm, which is consistent with the radius of $1\times 10^{11}$ cm derived by \citet{ChS09}, but much smaller than the estimate of Ts09 ($\sim8\times 10^{13}$ cm). Owing to large uncertainties in distance and reddening, and knowing that our one-zone model tends to overestimate the ejecta mass, we conclude that $M_\mathrm{ej}$ lies between 2 and 5 $M_\odot$, the $^{56}$Ni mass between 0.07 and 0.15 $M_\odot$, and the explosion energy between 1 and 6 foe. However, modelling of the early-time H$\alpha$ absorption restricts the total kinetic energy to be larger than 1.5 foe.

\section*{Acknowledgments}

We are grateful to the anonymous referee, whose comments helped to improve this work considerably.
Our thanks also go to the staff at the 3.58m Telescopio Nazionale Galileo (La Palma, Spain), the 2.2m Telescope of the Centro Astron\'{o}mico Hispano Alem\'{a}n (Calar Alto, Spain), the Asiago 1.22m and 1.82m Telescopes (Asiago, Italy) and the 1.08m AZT-24 telescope (Campo Imperatore, Italy).  ST acknowledges support by the Transregional Collaborative Research Centre TRR33 `The Dark Universe' of the German Research Foundation (DFG).  
We have made use of the NASA/IPAC Extragalactic Database (NED), operated by the Jet Propulsion Laboratory, California Institute of Technology, under contract with the National Aeronautics and Space Administration. This paper makes use of data obtained from the Isaac Newton Group Archive, which is maintained as part of the CASU Astronomical Data Centre at the Institute of Astronomy, Cambridge. 

\addcontentsline{toc}{chapter}{Bibliography}
\markboth{Bibliography}{Bibliography}
\bibliographystyle{mn2e}

\appendix

\section{Tables}

\begin{table*}
 \centering  
 \begin{minipage}{165mm}
  \caption{Optical and near-infrared magnitudes of the sequence stars in the field of NGC~4490.}
  \label{seq_mags}
  \begin{tabular}{@{}lcccccccc@{}}
  \hline
Star &	$U$	&	$B$	&	$V$	&	$R$	&	$I$	& $J$ & $H$ & $K$\\
  \hline
1	&$16.43\pm0.03$&$16.14\pm0.04$&$15.05\pm0.05$&$14.56\pm0.03$&$14.03\pm0.03$&$13.25\pm0.07$&$12.66\pm0.05$&$12.58\pm0.04$\\
2	&              &$18.59\pm0.05$&$17.59\pm0.04$&$16.95\pm0.03$&$16.42\pm0.03$&$15.61\pm0.06$&$15.02\pm0.03$&$14.97\pm0.05$\\
3	&              &$17.41\pm0.08$&$16.52\pm0.02$&$16.16\pm0.01$&$15.77\pm0.02$&    &  & \\
4	&              &$18.36\pm0.04$&$17.24\pm0.03$&$16.64\pm0.02$&$16.16\pm0.05$&    &  & \\
5	&              &$17.19\pm0.06$&$16.62\pm0.07$&$16.28\pm0.03$&$15.98\pm0.06$&    &  & \\
6	&              &$18.20\pm0.06$&$17.24\pm0.05$&$16.73\pm0.08$&$16.21\pm0.03$&    &  & \\
7	&              &$18.12\pm0.03$&$17.51\pm0.02$&$17.10\pm0.06$&$16.69\pm0.05$&    &  & \\
8	&$17.83\pm0.03$&$18.11\pm0.02$&$17.46\pm0.07$&$17.13\pm0.04$&$16.78\pm0.06$&    &  & \\
\hline\\
\\
\end{tabular}
\end{minipage}
\end{table*}

\begin{table*}
 \centering
 \begin{minipage}{137mm}
  \caption{$S$-correction added to the zero-point corrected SN
  magnitudes (\textit{instead of} colour-term corrections).}
  \label{Scorr}
  \begin{tabular}{@{}lcrrrrrrl@{}}
  \hline
   Date & JD & Phase\footnote{Phase in days with respect to the explosion date (JD = $2\,454\,528.80\pm0.15$). $B$-band maximum light occurred on day 18.3.} &$S_U$\quad\ & $S_B$\quad\ & $S_V$\quad\ & $S_R$\quad\ & $S_I$\quad\ & Instr.\footnote{CAFOS = Calar Alto 2.2m Telescope + CAFOS; DOLORES = 3.58m Telescopio Nazionale Galileo + DOLORES; AFOSC = Asiago 1.82m Copernico Telescope + AFOSC.} \\
   & $-2\,454\,000$ & (days) & & & & & & \\

\hline
08/03/08 & 533.59&   4.8 &              &$ 0.011\ \, $  &$-0.053\ \, $  &$ 0.010\ \, $  &$ 0.011$\footnote{Constant extrapolation.}     & CAFOS\\
12/03/08 & 537.53&   8.7 &              &$-0.057\ \, $  &$-0.002\ \, $  &$ 0.031\ \, $  &$ 0.013\ \, $  & AFOSC\\
14/03/08 & 539.76&  10.9 &$-0.042\ \, $ &$-0.003\ \, $  &$-0.035\ \, $  &$-0.006\ \, $  &$ 0.024\ \, $  & DOLORES\\
16/03/08 & 542.41&  13.6 &              &$ 0.004\ \, $  &$-0.043\ \, $  &$ 0.001\ \, $  &$ 0.025\ \, $  & CAFOS\\
28/03/08 & 554.47&  25.7 &$-0.105\ \, $ &               &$ 0.018\ \, $  &$ 0.031\ \, $  &$ 0.001\ \, $  & AFOSC\\
30/03/08 & 556.44&  27.6 &$-0.118\ \, $ &$-0.077\ \, $  &$ 0.033\ \, $  &$ 0.035\ \, $  &$ 0.018\ \, $  & AFOSC\\
03/04/08 & 559.50&  30.7 &$-0.138\ \, $ &$-0.102\ \, $  &$ 0.046\ \, $  &$ 0.042\ \, $  &$ 0.009\ \, $  & AFOSC\\
04/04/08 & 560.61&  31.8 &              &$ 0.058\ \, $  &$-0.095\ \, $  &$-0.002\ \, $  &$ 0.024\ \, $  & CAFOS\\
05/04/08 & 561.56&  32.8 &              &$ 0.059\ \, $  &$-0.096\ \, $  &$-0.002\ \, $  &$ 0.024\ \, $  & CAFOS\\
12/04/08 & 569.35&  40.6 &              &$ 0.066\ \, $  &$-0.107\ \, $  &$-0.003\ \, $  &$ 0.024\ \, $  & CAFOS\\
15/04/08 & 571.75&  42.9 &              &$ 0.005\ \, $  &$-0.081\ \, $  &$-0.012\ \, $  &$-0.025\ \, $  & DOLORES\\
25/04/08 & 582.35&  53.6 &              &$ 0.060\ \, $  &$-0.104\ \, $  &$-0.002\ \, $  &$ 0.025\ \, $  & CAFOS\\
01/05/08 & 588.47&  59.7 &              &$ 0.062\ \, $  &$-0.100\ \, $  &$ 0.001\ \, $  &$ 0.025\ \, $  & CAFOS\\
02/05/08 & 589.38&  60.6 &              &$-0.111\ \, $  &$ 0.037\ \, $  &$ 0.045\ \, $  &$-0.025\ \, $  & AFOSC\\
10/05/08 & 597.39&  68.6 &$-0.079\ \, $ &$ 0.005\ \, $  &$-0.065\ \, $  &$-0.013\ \, $  &$-0.037\ \, $  & DOLORES\\
11/06/08 & 629.48& 100.7 &$-0.018\ \, $ &$ 0.006\ \, $  &$-0.054\ \, $  &$-0.016\ \, $  &$-0.052\ \, $  & DOLORES\\
27/06/08 & 645.37& 116.6 &              &$ 0.063\ \, $  &$-0.078\ \, $  &$ 0.002\ \, $  &$-0.010\ \, $  & CAFOS\\
05/07/08 & 653.41& 124.6 &              &$ 0.064\ \, $  &$-0.078\ \, $  &$ 0.003\ \, $  &$-0.029\ \, $  & CAFOS\\ 
24/07/08 & 672.39& 143.6 &$-0.067^c$    &$-0.067\ \, $  &$ 0.015\ \, $  &$ 0.050\ \, $  &$-0.043\ \, $  & AFOSC\\
24/11/08 & 794.69& 265.9 &              &$ 0.008^c$     &$-0.177\ \, $  &$ 0.026\ \, $  &$-0.045\ \, $  & DOLORES\\
22/12/08 & 822.76& 294.0 &              &$ 0.008^c$     &$-0.190^c$     &$ 0.033\ \, $  &$-0.058^c$     & DOLORES\\
19/02/09 & 881.54& 352.7 &              &$ 0.065^c$     &$-0.216^c$     &$-0.001\ \, $  &$-0.510^c$     & CAFOS\\
\hline
\end{tabular}
\end{minipage}
\end{table*}

\label{lastpage}

\end{document}